\def\apj{ApJ}

\def\apjs{ApJS}
\def\mnras{MNRAS}
\def\aa{A\&A}

\def\aj{AJ}

\documentclass[useAMS,usenatbib]{mn2e}
\usepackage{graphicx}
\usepackage{url}

\title[Binary star modeling of galactic SEDs]{Integrated spectral energy distributions of binary star composite stellar populations}
\author[Z.~M. Li et al.]{Zhongmu Li$^{1,2}$ \thanks{E-mail:
zhongmu.li@gmail.com}, Liyun Zhang$^{3}$, Jinzhong Liu$^{4}$\\
$^{1}$Institute for Astronomy and History of Science and Technology, Dali
University, Dali 671003, China\\
$^{2}$Key Laboratory of Optical Astronomy, National Astronomical Observatories, Chinese Academy of
Sciences, Beijing 100012, China\\
$^{3}$College of Science/Department of Physics\&NAOC-GZU-Sponsored Center for Astronomy Research, Guizhou University,
Guiyang 550025, China\\
$^{4}$Xinjiang Astronomical Observatory, Chinese Academy of Sciences, Urumqi 830011, China}

\begin{document}

\date{Accepted 1988 December 15. Received 2011 December 14; in original form 1988 October 11}

\pagerange{\pageref{firstpage}--\pageref{lastpage}} \pubyear{2011}

\maketitle

\label{firstpage}

\begin{abstract}
This paper presents theoretical integrated spectral energy
distributions (SEDs) of binary star composite stellar populations
(bsCSPs) in early-type galaxies, and how the bsCSP model can be used
for spectral studies of galaxies. All bsCSPs are built basing on
three adjustable inputs (metallicity, ages of old and young
components). The effects of binary interactions and stellar
population mixture are taken into account. The results show some
UV-upturn SEDs naturally for bsCSPs. The SEDs of bsCSPs are affected
obviously by all of three stellar population parameters, and the
effects of three parameters are degenerate. This suggests that the
effects of metallicity, and the ages of the old (major in stellar
mass) and young (minor) components of stellar populations should be
taken into account in SED studies of early-type galaxies. The
sensitivities of SEDs at different wavelengths to the inputs of a
stellar population model are also investigated. It is shown that UV
SEDs are sensitive to all of three stellar population parameters,
rather than to only stellar age. Special wavelength ranges according
to some SED features that are relatively sensitive to stellar
metallicity, young-component age, and old-component age of bsCSPs
are found by this work. For example, the shapes of SEDs with
wavelength ranges of 5110-5250\,$\rm \AA$, 5250--5310\,$\rm \AA$,
5310--5350\,$\rm \AA$, 5830--5970\,$\rm \AA$, 20950--23550\,$\rm
\AA$ are relatively sensitive to the stellar metallicity of bsCSPs.
The shapes of SEDs within 965-985\,$\rm \AA$, 1005--1055\,$\rm \AA$,
1205--1245\,$\rm \AA$ are sensitive to old-component age, while SED
features within the wavelength ranges of 2185--2245\,$\rm \AA$,
2455--2505\,$\rm \AA$, 2505--2555\,$\rm \AA$, 2775--2825\,$\rm \AA$,
2825--2875\,$\rm \AA$ to young-component age. The results suggest
that some line indices within these special wavelength ranges are
possibly better for stellar population studies compared to the
others, and greater weights may be given to these special SED parts
in the determination of the stellar population parameters of
early-type galaxies from fitting SEDs via bsCSPs.
\end{abstract}


\section{Introduction}
Integrated spectral energy distribution (SED) is one of the most
important tools for studying early-type galaxies. It can be used to
investigate a lot of physical information of galaxies, because SEDs
at different wavelengths are usually dominated by different physical
parameters. For example, SED fitting can give insight to stellar
mass, age, metallicity, and redshift of galaxies. On studies of SEDs
of galaxies, one can use both observational spectral templates or
theoretical SEDs. When theoretical SEDs are taken for studying the
stellar populations of early-type galaxies, many evolutionary
stellar population synthesis models, e.g., \citet{Leitherer:1999}
and \citet{Vazquez:2007} (Starburst99), \citet{Vazdekis1999},
\citet{Schulz:2002}, \citet{Cervino:2002}, \citet{Robert:2003},
\citet{Bruzual:2003} (BC03), \citet{LeBorgne:2004} (PEGASE-HR),
\citet{Maraston:2005} (M05), \citet{Lanccon:2008},
\citet{Molla:2009}, are usually used. However, all of these models
are single star simple stellar population (ssSSP) models, which do
not take the effects of binary evolution and population mixture into
account. Meanwhile, observations in the Galaxy show that more than a
half of stars are in binaries. Studies on the evolution of binary
stars show that binary stars have significantly different evolution
processes comparing to single stars. When binary stars are used to
model the SEDs of stellar populations of early-type galaxies, some
recent studies show that binary evolution can affect number or
spectral population synthesis studies significantly
\citep{VanBever:1999,VanBever:2003,
Li:2008MNRAS,Li:2008ApJ,Li:2009RAA,Li:2010RAA,Zhang:2010MNRAS,Han:2010IAUS},
especially in the UV (e.g., \citealt{Belkus:2003}), X-ray
\citep{VanBever:2000}, and optical bands \citep{Han:2007}.
Therefore, it is necessary to model the SEDs of galaxies via binary
stars. In addition, a lot of works (e.g., \citealt{Marmol:2009})
showed that galaxies, even early-type ones, should undergo more than
one star bursts. It means that there are multiple stellar
populations in early-type galaxies. This calls for composite stellar
population (CSP) models instead of simple stellar population (SSP)
models to give more detailed SED studies to early-type galaxies.
\citet{Han:2007} studied the UV-upturn of SEDs of elliptical
galaxies via binary stars, but all populations are assumed the solar
metallicity ($Z$ = 0.02). Because many works show wide metallicity
ranges for early-type galaxies, it is necessary to study the SEDs of
early-type galaxies using a more advanced binary star composite
stellar population (bsCSP) model, which covers a large metallicity
range. This paper presents a new model, via assuming an old (major
in stellar mass) and a young (minor) SSP component for a bsCSP. The
assumption of two populations for a bsCSP is taken because
early-type galaxies have relatively simple star formation histories
comparing to other galaxies, and it is possible to give the main
information of stellar populations of early type-galaxies via two
stellar population components. Our work aims to investigate how
model inputs affect the SEDs of populations and to find the
sensitivities of different parts of SEDs to model inputs.

The paper is organized as follows. In Sect. 2, we introduce the
binary star composite stellar population model. In Sect. 3, we study
how different inputs of a bsCSP model affect the SEDs of
populations. In Sect. 4, we study the sensitivities of SEDs at
different wavelengths to three main parameters of a bsCSP. In Sect.
5, the bsCSP model is used to determine some important parameters of
three elliptical galaxies. Finally, we give our conclusion and
discussion in Sect. 6.

\section{Binary star composite stellar population model}
There are two steps to build the SEDs of bsCSPs. First, a large
database of SEDs of binary star simple stellar populations (bsSSPs)
is built. Stars in a bsSSP are assumed to form at a star burst and
have the same metallicity. Second, the SEDs of bsCSPs are built on
the basis of the SED database of bsSSPs. Because there is no common
results for the stellar mixture of galaxies, we take a simple way to
model bsCSPs. Each bsCSP is assumed to contain a pair of old and
young components with the same metallicity. This assumption is in
agreement of previous studies that early-type galaxies are dominated
by old populations and there is only a little fraction of young
population in such galaxies. The mass fraction of young component is
assumed to be dependent on the ages of two components of bsCSPs,
which is calculated by formula (1). This means that the mass
fraction of young component declines exponentially with increasing
difference between the ages of old and young components. This is in
agreement with some previous studies on the star formation histories
of early-type galaxies, e.g., \citet{Thomas:2005}.
\begin{equation}
  F_{\rm 2} = 0.5 ~{\rm exp}[\frac{t_{\rm2}-t_{\rm1}}{\rm
  \tau}]
\end{equation}
where $F_{\rm 2}$ is the mass fraction of young component;
$t_{\rm1}$ and $t_{\rm2}$ are the ages of the old and young
components of a bsCSP. As a standard model, $\tau$ is taken as 3.02,
according to the observational fraction of bright early type
galaxies with recent ($\leq$ 1\,Gyr) star formation at a level more
than 1\%--2\% \citep{Yi:2005,Li:2007AA}.

The important ingredients and specialties of our model are as
follows. An initial mass function (IMF) of \citet{Chabrier:2003}
with lower and upper mass limits of 0.1 and 100 M$_\odot$
respectively is taken for bsCSPs. The evolution of binaries is
calculated via the rapid stellar evolution code of
\citet{Hurley:2002} (hereafter Hurley code), so that most of binary
evolution processes such as mass transfer, mass accretion,
common-envelope evolution, collisions, supernova kicks and angular
momentum loss are included. Different mass transfer mechanisms,
i.e., dynamical mass transfer, nuclear mass transfer and thermal
mass transfer are taken into account using the results of many works
(e.g., \citealt{Tout:1997,Hjellming:1987}). One can see
\citealt{Hurley:2002} for more details. According to a previous
work, in all binary interactions, Roche lobe overflow (RLOF) and and
common envelope (CE) may affect the formation of hot subdwarfs and
then the UV spectra of stellar populations more significantly
\citep{Han:2007}.

The default values in Hurley code, i.e., 0.5, 1.5, 1.0, 0.0, 0.001,
3.0, 190.0, 0.5, and 0.5, are taken for wind velocity factor
($\beta_{\rm w}$), Bondi-Hoyle wind accretion faction ($\alpha_{\rm
w}$), wind accretion efficiency factor ($\mu_{\rm w}$), binary
enhanced mass loss parameter ($B_{\rm w}$), fraction of accreted
material retained in supernova eruption ($\epsilon$),
common-envelope efficiency ($\alpha_{\rm CE}$), dispersion in the
Maxwellian distribution for the supernovae kick speed ($\sigma_{\rm
k}$), Reimers coefficient for mass loss ($\eta$), and binding energy
factor ($\lambda$), respectively. Although the above default values
remain somewhat large uncertainties, the results for spectral
stellar population synthesis will be not affected too much by the
uncertainties of these parameters. We test the effects of
uncertainties in these parameters by comparing four modes with
different input values. For convenience, the mode of stellar
evolution by taking the default values will be cited as default mode
hereafter. Then the evolutionary results of other three modes (A, B
and C) are compared to that of the default mode. In mode A, we take
the lowest values for all parameters (i.e., 0.125, 0.0, 0.0, 0.0,
-1.0, 0.5, 0.0, 0.0 for $\beta_{\rm w}$, $\alpha_{\rm w}$, $\mu_{\rm
w}$, $B_{\rm w}$, $\epsilon$, $\alpha_{\rm CE}$, $\sigma_{\rm k}$
and $\eta$, respectively) to evolve 5000 binaries with metallicities
($Z$) of 0.01, 0.02, and 0.03 from zero-age main sequence (ZAMS) to
15\,Gyr, with an age interval of 0.1\,Gyr. Then we compared the
effective temperature ($T_{\rm eff}$) and surface gravity (log $g$)
of stars in mode A to those obtained in the default mode. It shows
that the two main stellar parameters ($T_{\rm eff}$ and log $g$)
change only 0.20\% on average when taking mode A instead of default
mode. Then in mode B, similar to mode A, but the largest or large
(for $\sigma_{\rm k}$) values are taken for all parameters (i.e.,
7.0, 2.0, 1.0, 1.0$^6$, 1.0, 10.0, 570.0, 2.0 for $\beta_{\rm w}$,
$\alpha_{\rm w}$, $\mu_{\rm w}$, $B_{\rm w}$, $\epsilon$,
$\alpha_{\rm CE}$, $\sigma_{\rm k}$ and $\eta$, respectively) when
evolving binaries. It shows that the average change of stellar
parameters is 0.48\% compared to the default mode, which is larger
than in mode A. In the third mode, i.e., mode C, we set some random
values to evolve 5000 binaries. The values for $\beta_{\rm w}$,
$\alpha_{\rm w}$, $\mu_{\rm w}$, $B_{\rm w}$, $\epsilon$,
$\alpha_{\rm CE}$, $\sigma_{\rm k}$, and $\eta$ are set as 3.0, 1.0,
0.3, 100.0, 0.5, 5.0, 300.0 and 0.7, respectively. This mode shows
an average change in stellar parameters of 0.35\% compared to the
default mode. Because the fitted formulae used by Hurley code to
evolve stars can lead to uncertainties up to about 5\%
\citep{Hurley:2002}, the changes of stellar evolutionary results
that are caused by the uncertainties in the default parameters, are
much less than the uncertainties result from using fitted formulae
to evolve stars. Therefore, we conclude that the uncertainties in
the default values of Hurley code will not change the main results
of statistical studies like spectral stellar population synthesis.
Thus we take the default values of Hurley code to evolve all stars
in this work. Note that 3.0 is taken for $\alpha_{\rm CE}$ here also
because it can produce enough double-degenerates while a widely used
value of 1.0 can not when \cite{Hurley:2002} tried to check the
effects of input parameters of Hurley code via population synthesis
of birth rates and Galactic numbers of the various types of binary.

When generating the star sample of populations, an uniform
distribution is taken for the ratio ($q$, 0--1) of the mass of the
secondary to that of the primary (\citealt{Mazeh:1992};
\citealt{Goldberg:1994}), and the mass of the secondary is then
calculated from $q$ and the primary mass, where the mass of the
primary is generated via a Monte Carlo technique according to the
selected IMF. The separation ($a$) of two components of a binary is
generated following the assumption that the fraction of binary in an
interval of log($a$) is constant when $a$ is big (10$R_\odot$ $< a
<$ 5.75 $\times$ 10$^{\rm 6}$$R_\odot$) and it falls off smoothly
when $a$ is small ($\leq$ 10$R_\odot$) \citep{Han:1995}, which can
be written as
\begin{equation}
  a~.p(a) = \left\{
            \begin{array}{ll}
            a_{\rm sep}(a/a_{\rm 0})^{\psi}, &~a \leq a_{\rm 0}\\
            a_{\rm sep}, &~a_{\rm 0} < a < a_{\rm 1}\\
     \end{array}
    \right.
\end{equation}
where $a_{\rm sep} \approx 0.070, a_{\rm 0} = 10R_{\odot}, a_{\rm 1}
= 5.75 \times 10^{\rm 6}R_\odot$ and $\psi \approx 1.2$. An uniform
distribution is taken for the eccentricity ($e$) of each binary
system. Most binary interactions, e.g., mass transfer,
common-envelope evolution, collisions, and tidal interactions, are
taken into account in the bsCSP model. The fraction of binary stars
in each population (4\,000\,000 single stars) is taken as the
typical value of the Galaxy, i.e., 50\%. When calculating the SEDs
and of stellar populations, the BaSeL 3.1 spectral library
\citep{Lejeune:1997, Lejeune:1998, Westera:2002} is used because of
its wide wavelength coverage and reliability. On average, the
spectral library leads to about 3\% uncertainties in the final SEDs
of bsCSPs.

\section{Effects of model inputs on SEDs}
\subsection{Effect of stellar metallicity}
Most previous works took the solar metallicity for studying the SEDs
of early-type galaxies, especially elliptical galaxies, but stellar
metallicities ($Z$) of early-type galaxies distribute in a large
range (e.g., 0.005 -- 0.05, \citealt{Gallazzi:2005}). Therefore, it
is interesting to investigate how stellar metallicity changes the
SEDs of stellar populations of early-type galaxies. The effects of
metallicity on the SEDs of bsCSPs are investigated in this section.
Figs. 1 and 2 show some results. In each panel of the two figures,
the SEDs of bsCSPs with different metallicities but the same pair of
old and young-component ages are compared. Symbols `$t_{\rm 1}$' and
`$t_{\rm 2}$' denote the ages of old and young-components,
respectively. As can be seen, stellar metallicity affects the SEDs
of bsCSPs significantly. In particular, UV SED is shown to be very
sensitive to stellar metallicity. When stellar metallicity changing
from 0.004 to 0.03, the logarithmic flux in UV band can change as
large as 1.8\,dex. Even stellar metallicity changes only from 0.2 to
0.01, the UV spectra of a bsCSPs can change a lot for most of
stellar populations. This suggests that stellar metallicity should
be taken into account when explaining the UV spectra of early-type
galaxies. In addition, the result shows no well-regulated
correlation between metallicity and the change of SEDs of bsCSPs,
comparing to a solar metallicity case. Therefore, it is useful to
use bsCSPs with metallicities in a wide range to study the SEDs of
early-type galaxies.

\begin{figure} 
\centering
\includegraphics[angle=-90,width=0.35\textwidth]{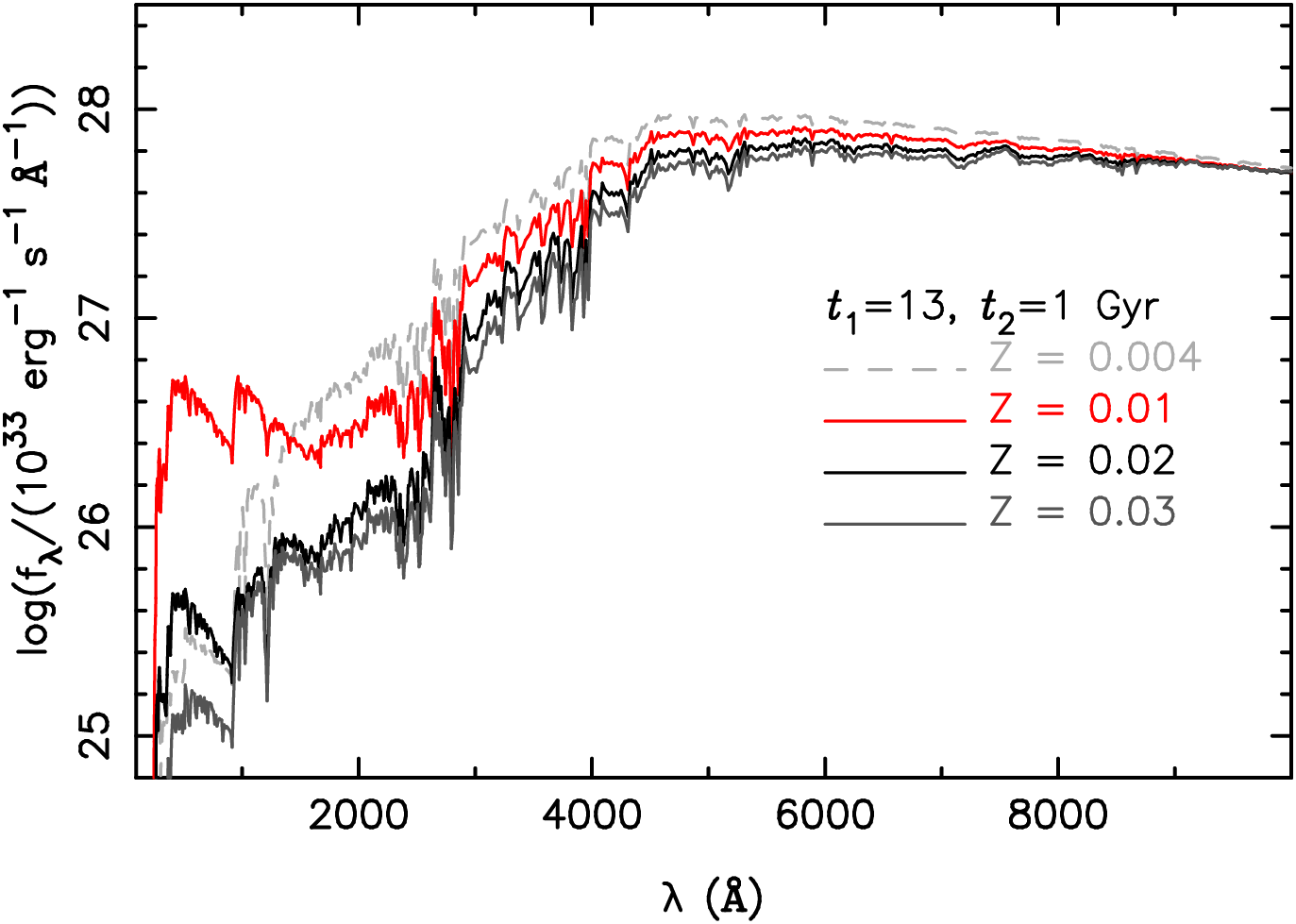}
\includegraphics[angle=-90,width=0.35\textwidth]{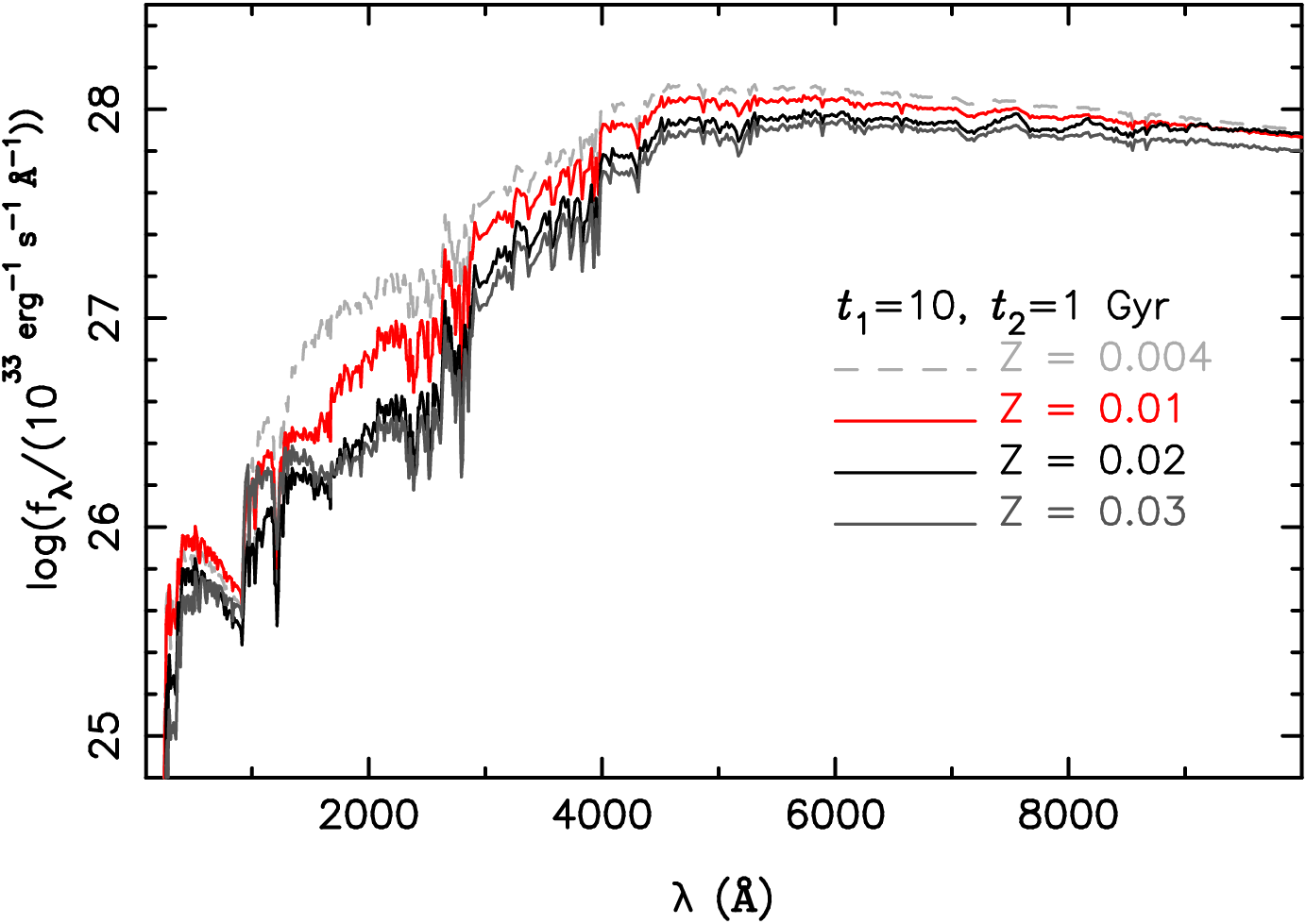}
\includegraphics[angle=-90,width=0.35\textwidth]{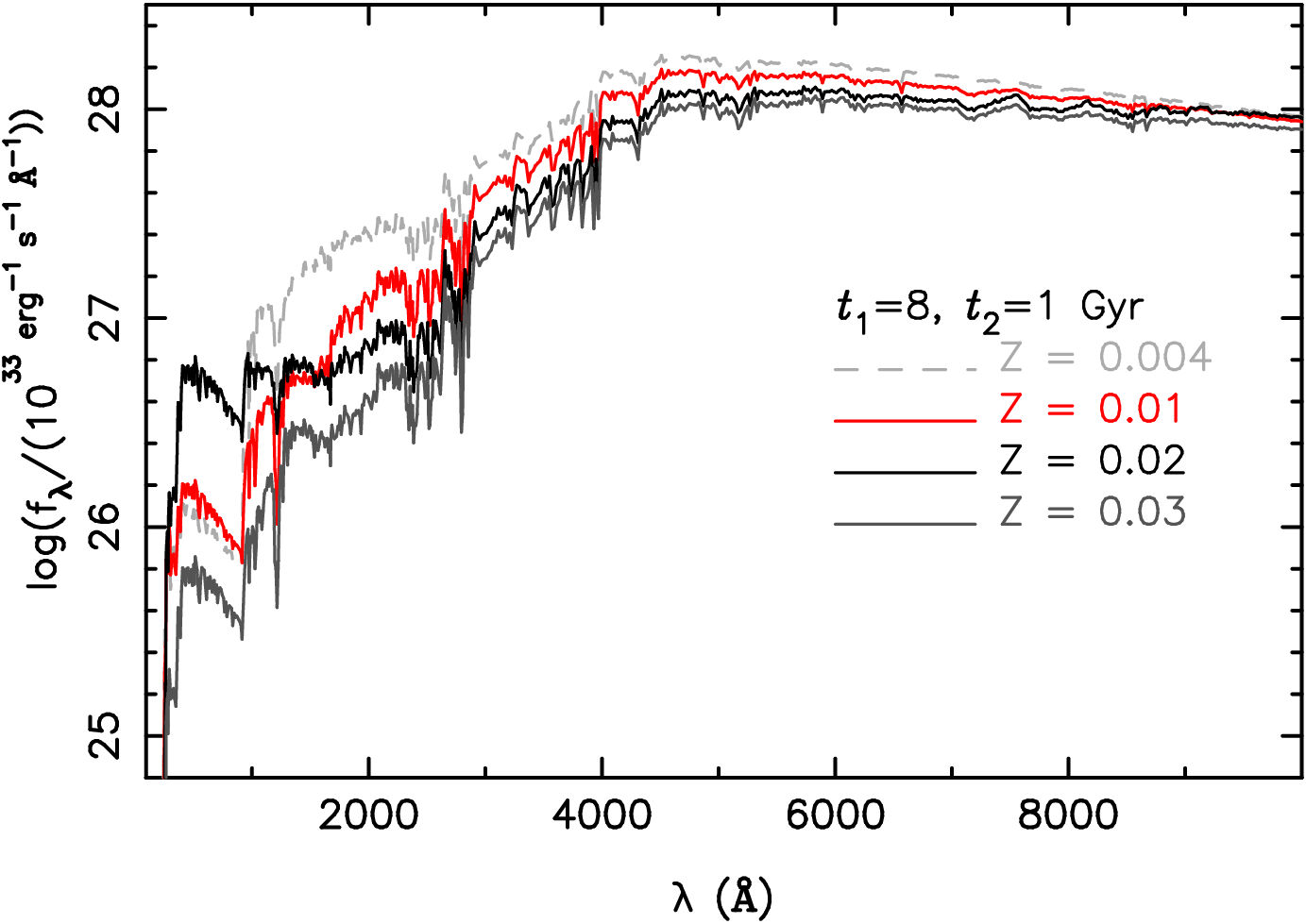}
\includegraphics[angle=-90,width=0.35\textwidth]{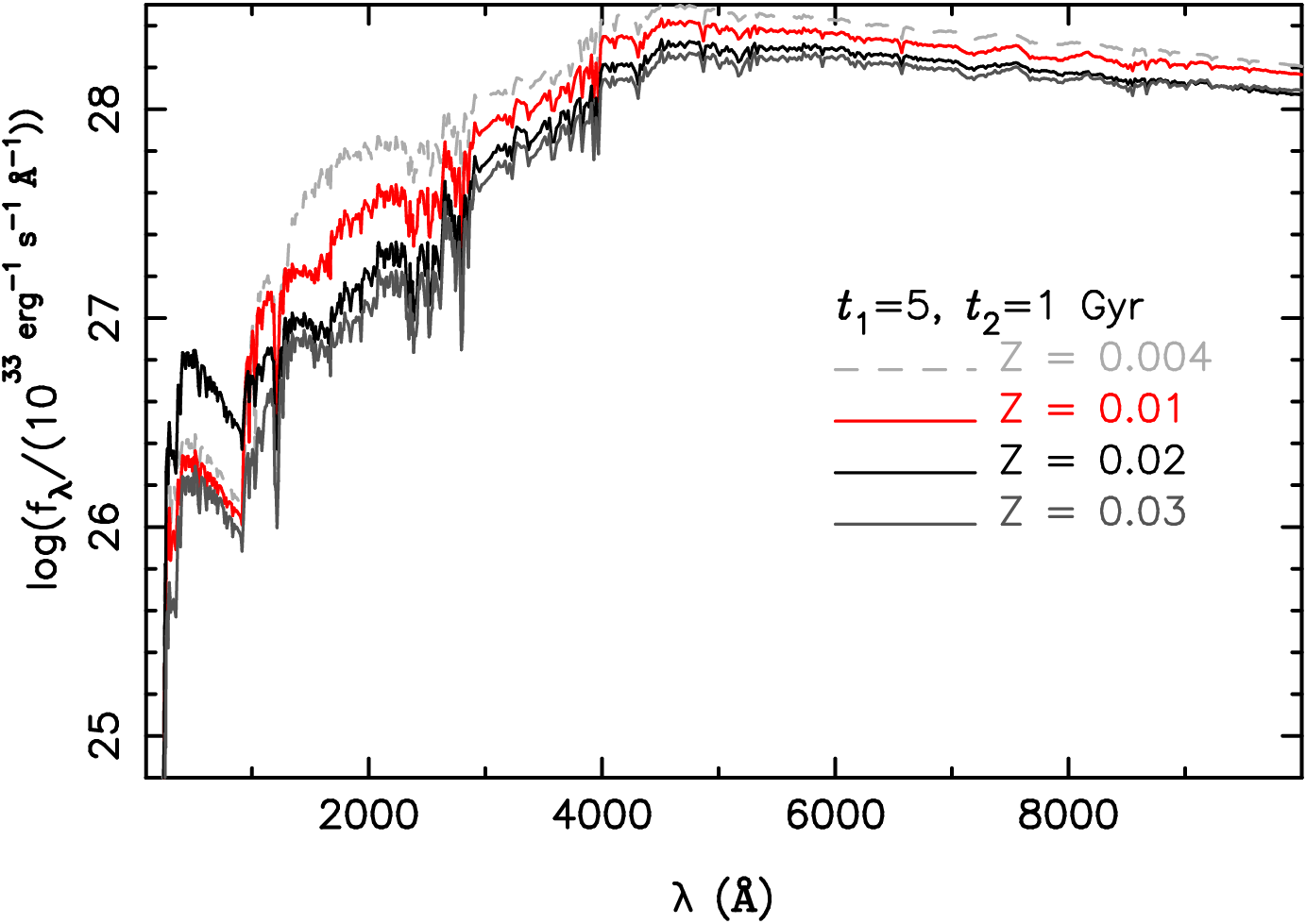}
\caption{Comparison of SEDs of bsCSPs with different stellar metallicities.
All bsCSPs have the same age for their young components, i.e., $t_{\rm 2}$ = 1\,Gyr.
The ages of old-components of bsCSPs, $t_{\rm 1}$, are set to be different from 13 to 5\,Gyr in each panel.
Stellar populations with different metallicities are shown in various colours as the remarks in each panel.}
\end{figure}

\begin{figure} 
\centering
\includegraphics[angle=-90,width=0.35\textwidth]{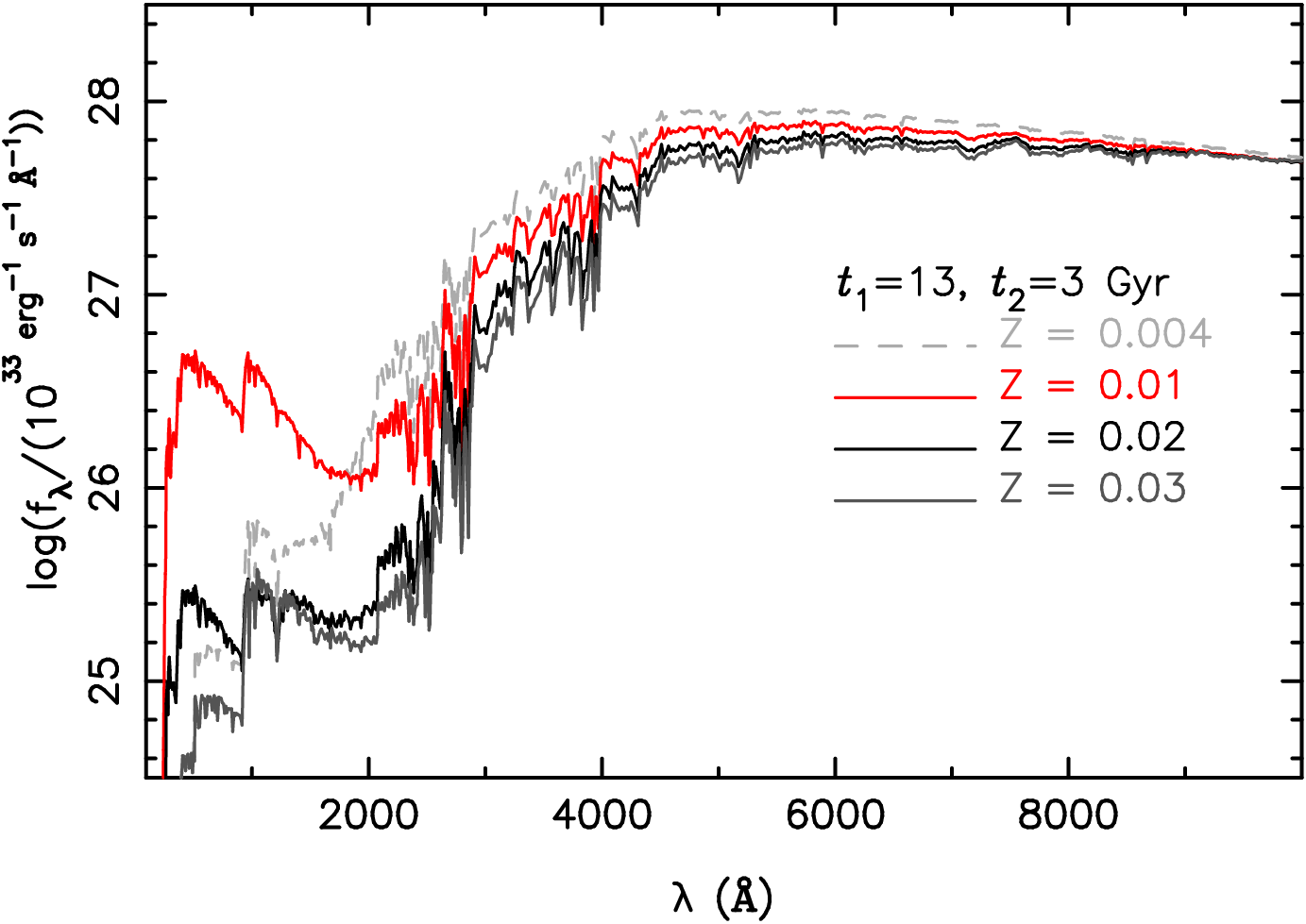}
\includegraphics[angle=-90,width=0.35\textwidth]{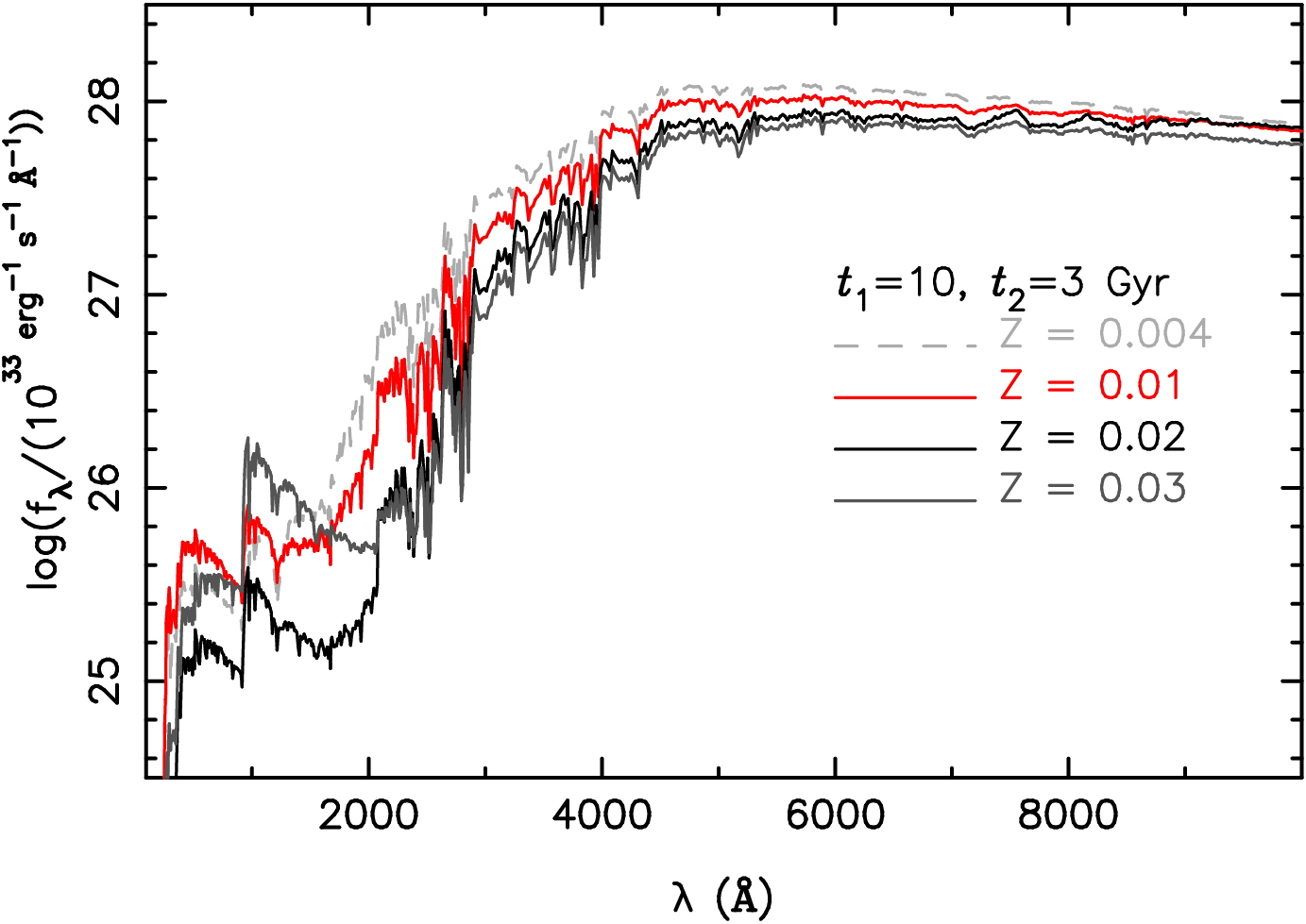}
\includegraphics[angle=-90,width=0.35\textwidth]{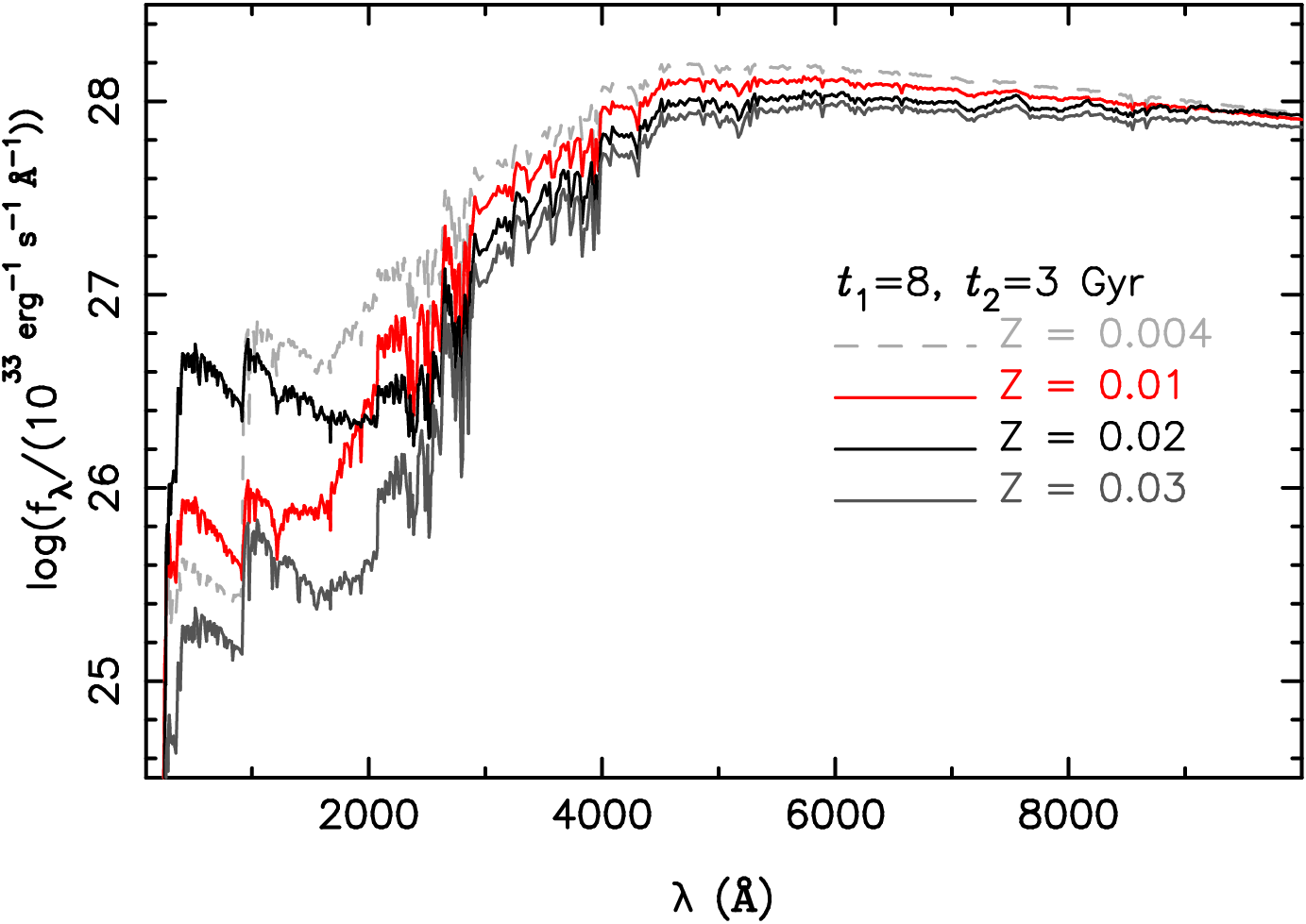}
\includegraphics[angle=-90,width=0.35\textwidth]{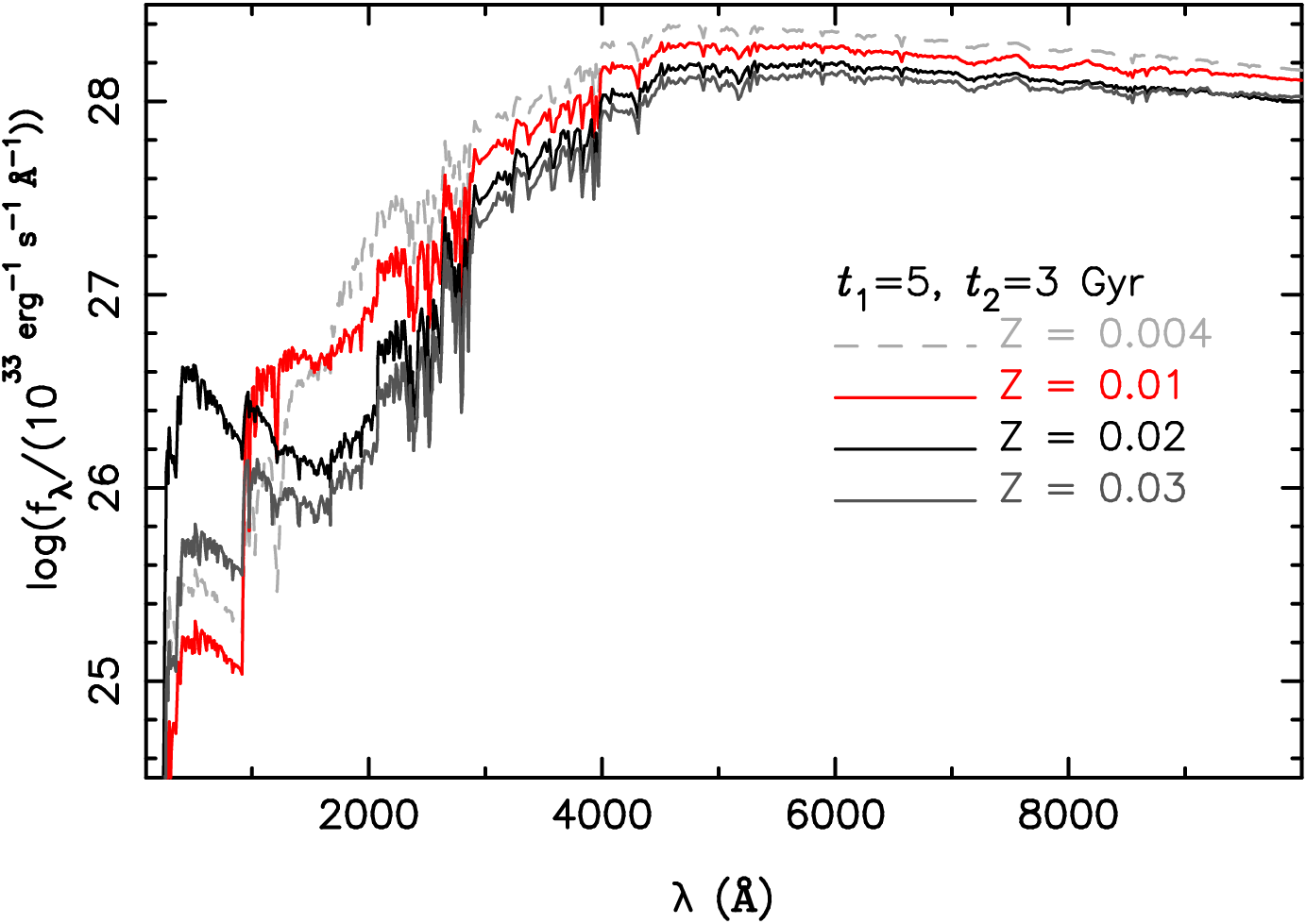}
\caption{Similar to Fig. 1, but for bsCSPs with a young-component age of 3\,Gyr.}
\end{figure}

\subsection{Effect of old-component age}
This section investigates how the age of old component, $t_{\rm 1}$,
affects the SED of bsCSPs. As examples, some stellar populations
with metallicities of 0.01, 0.02, and 0.003 are investigated. The
ages of young components of bsCSPs are taken as 1, 2, 3, or 5\,Gyr,
and the ages of old components differ from 13 to 5\,Gyr. The main
results are shown in Figs. 3 to 5. It is found that the age of old
component affects the SEDs of bsCSPs obviously. The optical and
near-infrared flux decreases with increasing old-component age
($t_{\rm 1}$), for fixed metallicity ($Z$) and young-component age
(t$_{\rm 2}$). However, the flux and shape of UV-band SEDs of bsCSPs
do not change with $t_{\rm 1}$ regularly. It seems that UV SED is
not only sensitive to the old-component age of bsCSPs.

\begin{figure} 
\centering
\includegraphics[angle=-90,width=0.35\textwidth]{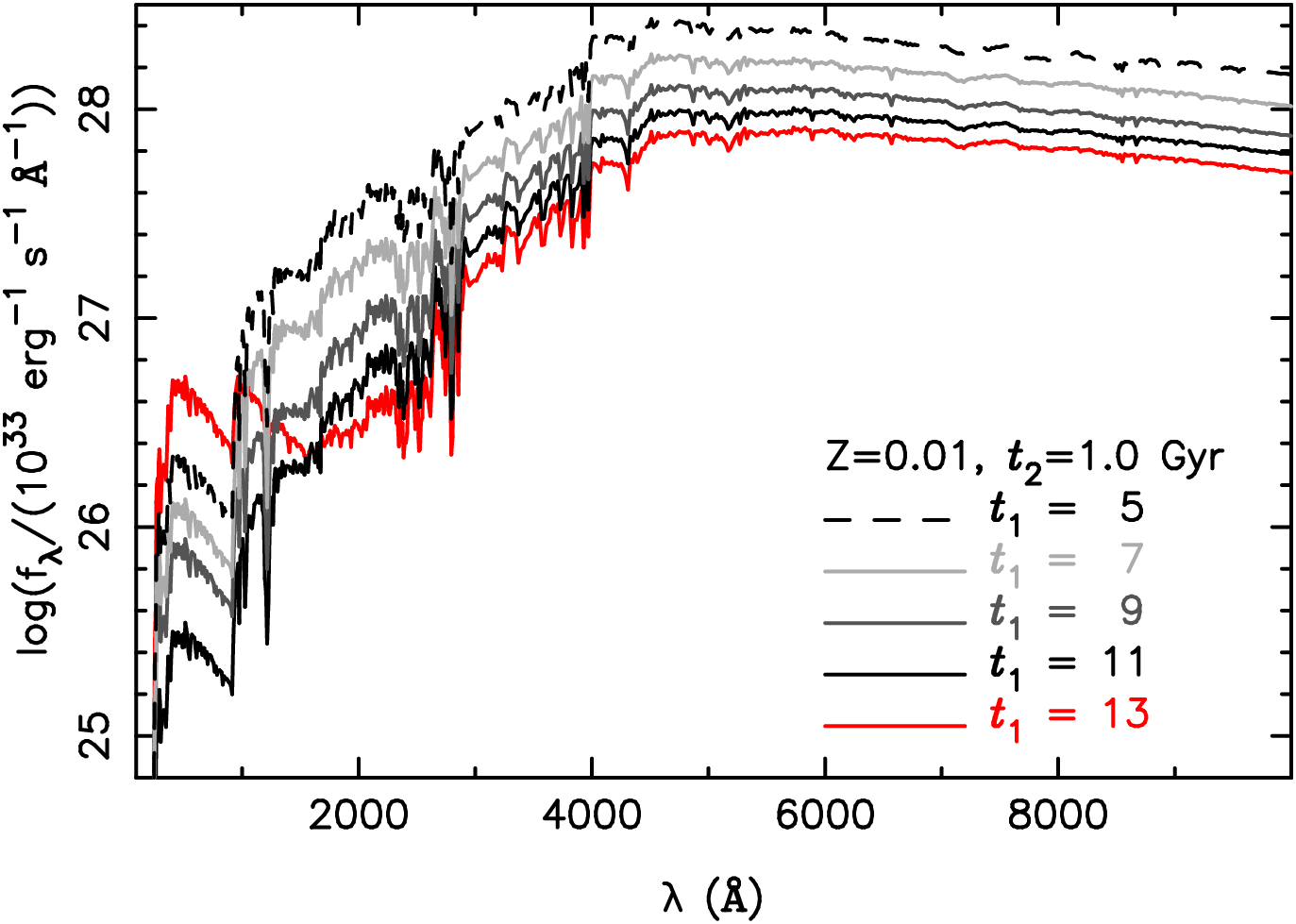}
\includegraphics[angle=-90,width=0.35\textwidth]{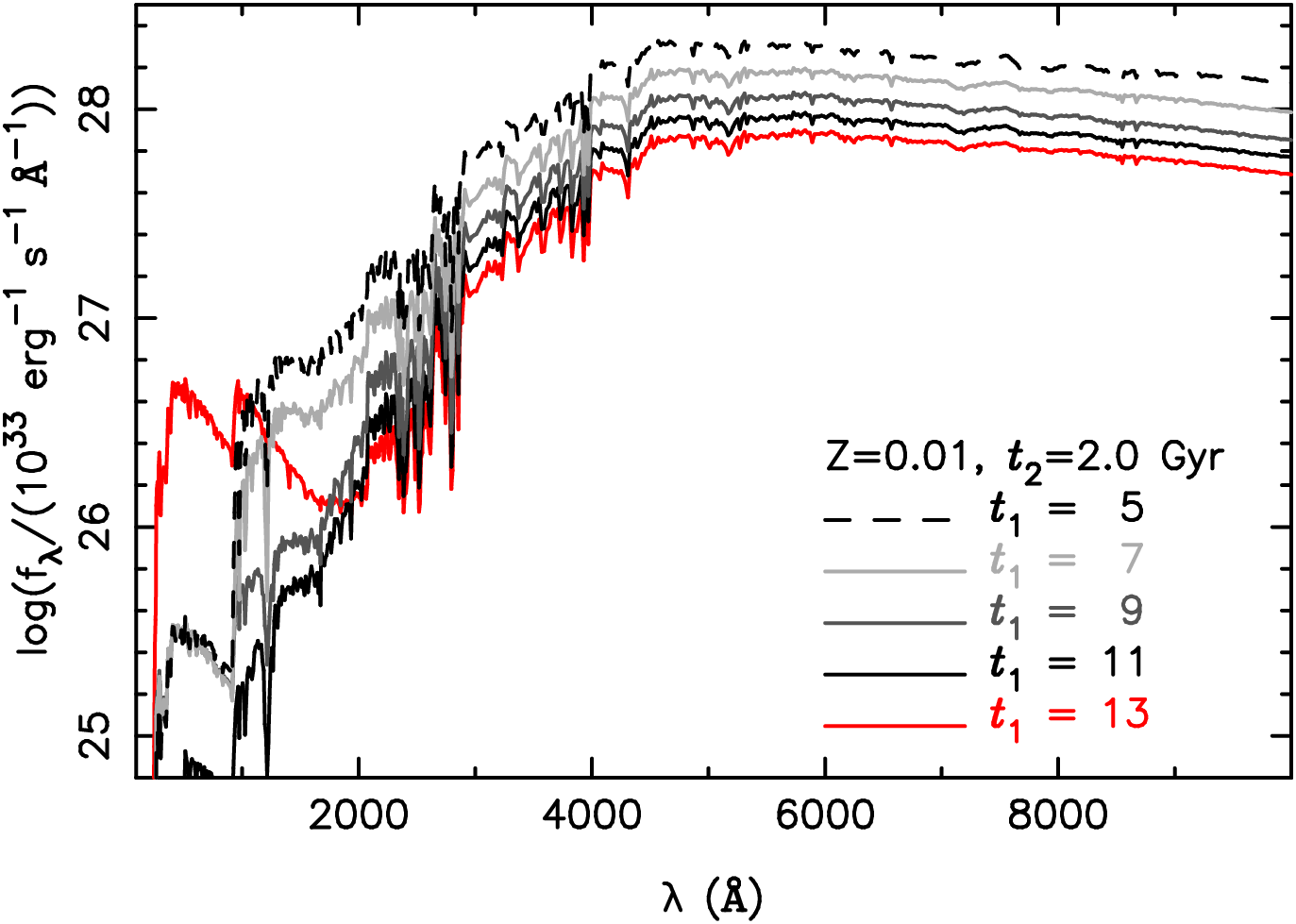}
\includegraphics[angle=-90,width=0.35\textwidth]{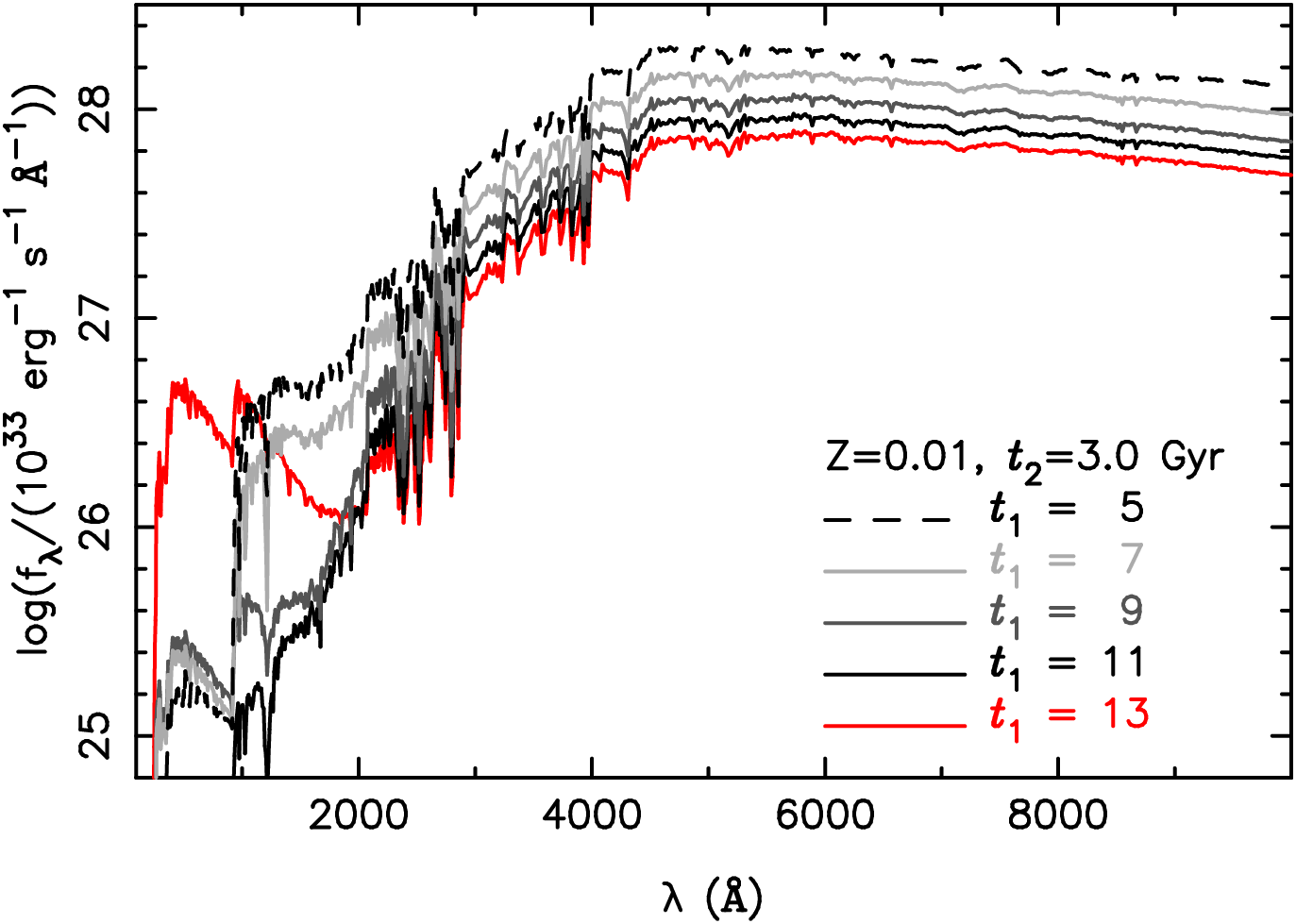}
\includegraphics[angle=-90,width=0.35\textwidth]{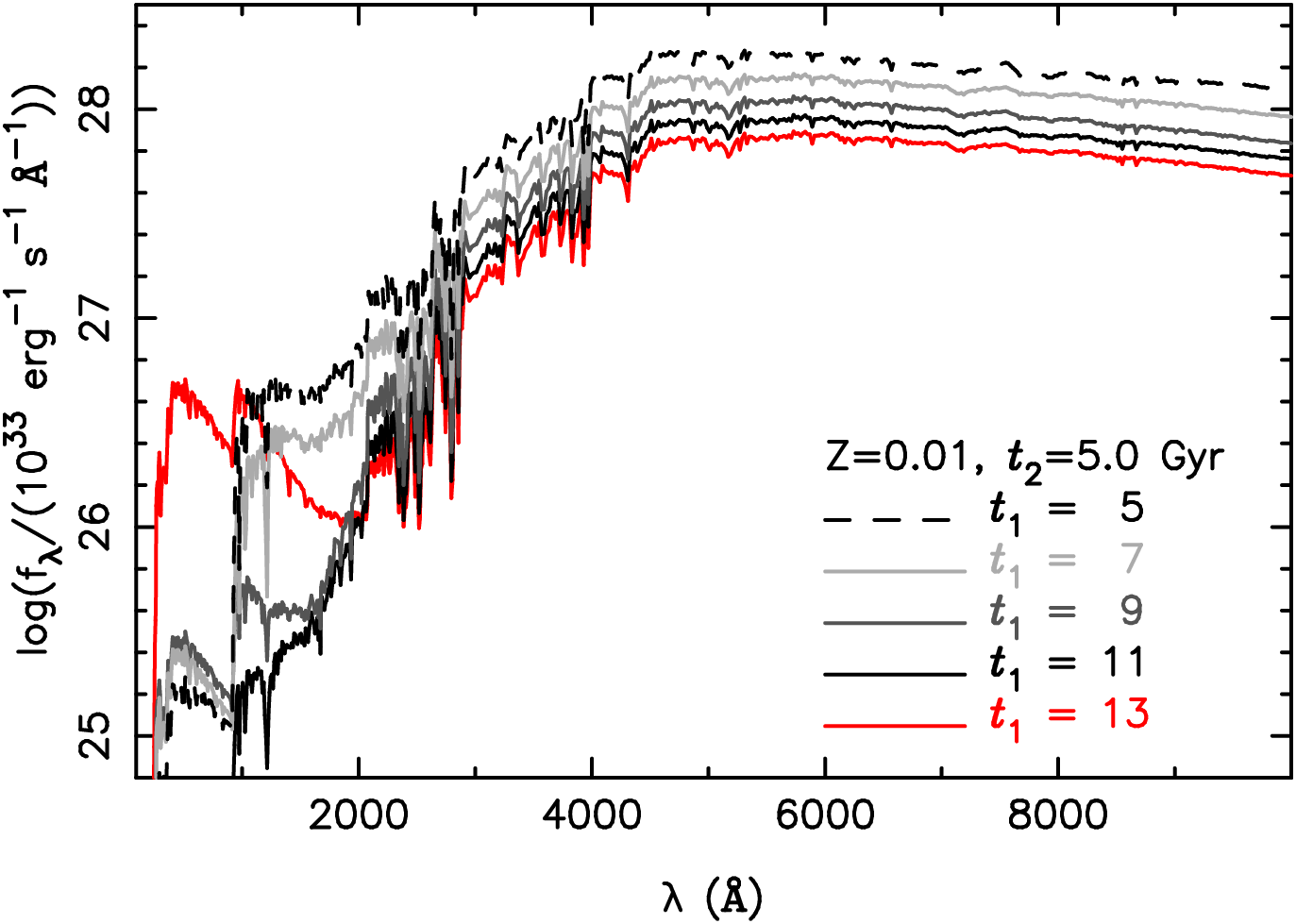}
\caption{Comparison of SEDs of bsCSPs with different old-component ages.
The ages of old-components of bsCSPs, $t_{\rm 2}$, are set to be different (1, 2, 3, 5\,Gyr) in four panels,
while the stellar metallicity, $Z$, is set to be the same.
Stellar populations with different old-component ages are shown in various colours or line types.}
\end{figure}

\begin{figure} 
\centering
\includegraphics[angle=-90,width=0.35\textwidth]{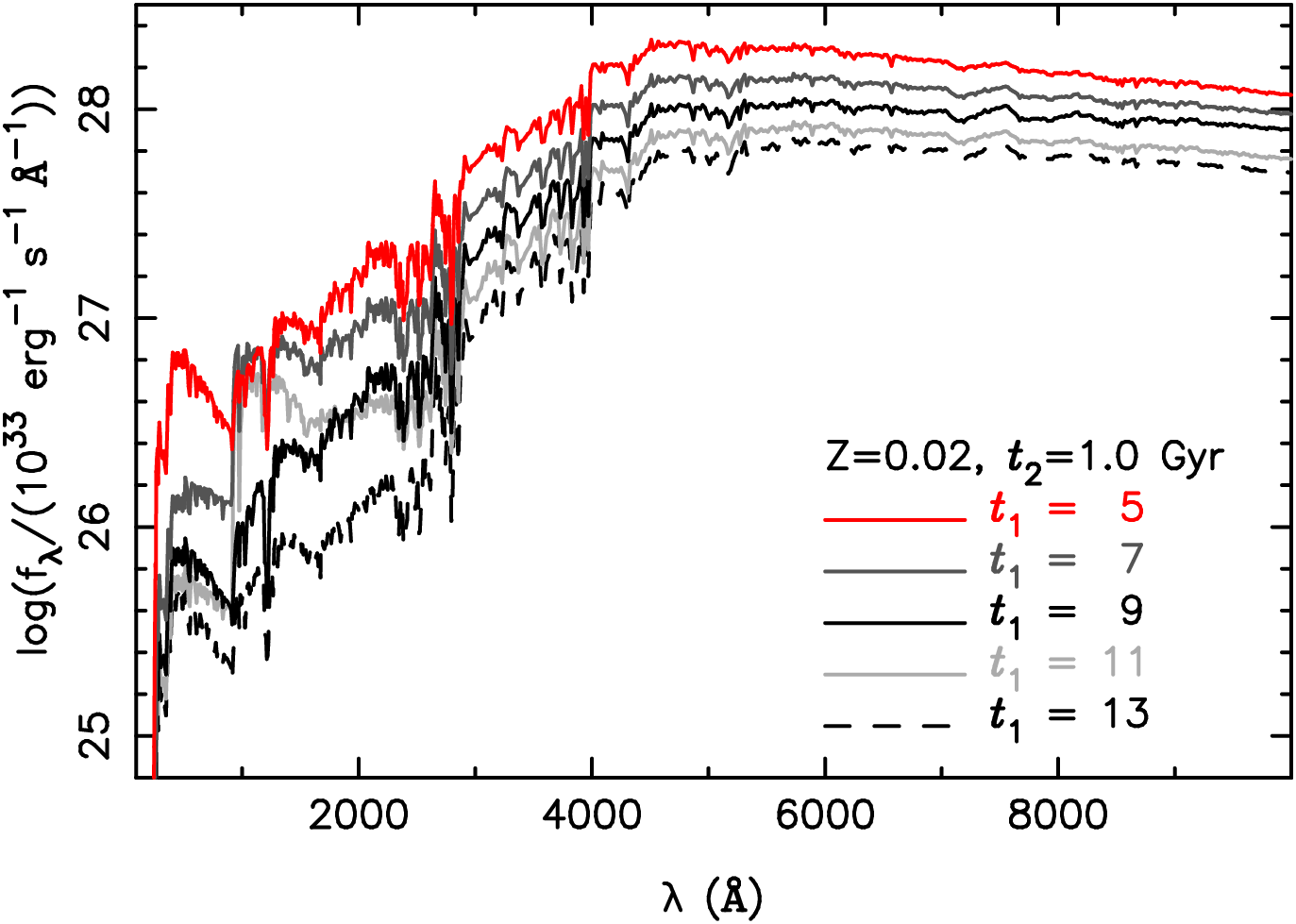}
\includegraphics[angle=-90,width=0.35\textwidth]{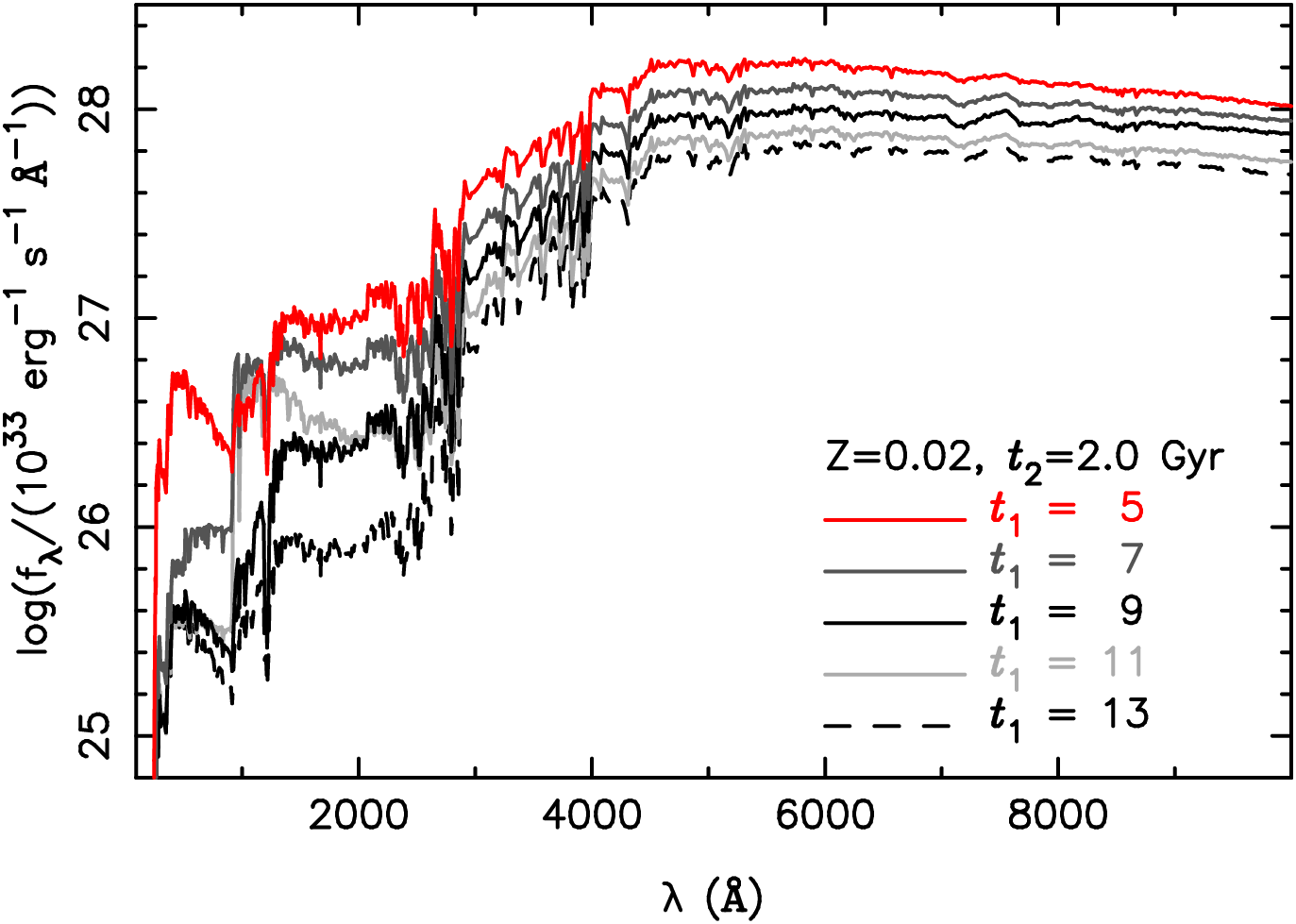}
\includegraphics[angle=-90,width=0.35\textwidth]{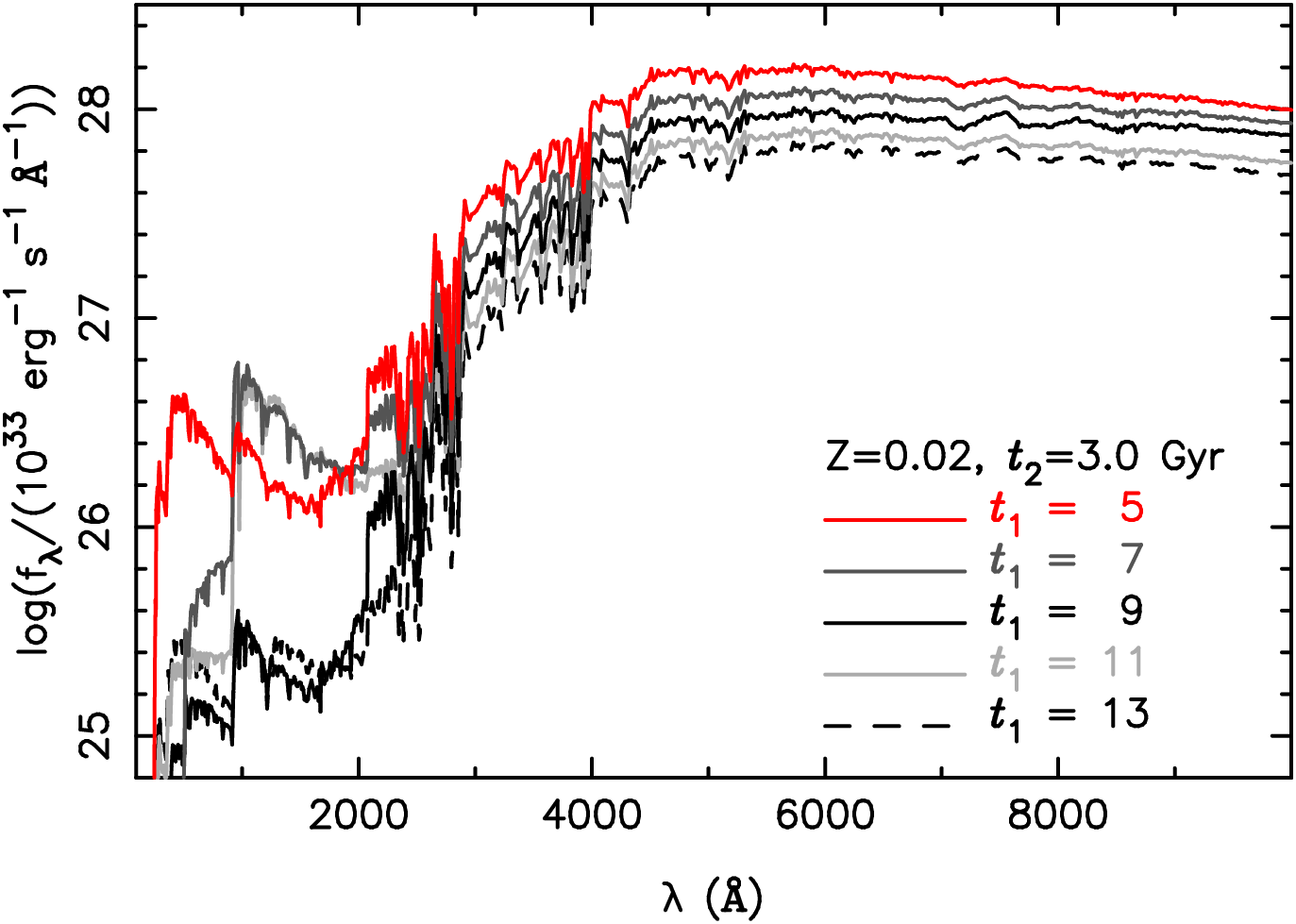}
\includegraphics[angle=-90,width=0.35\textwidth]{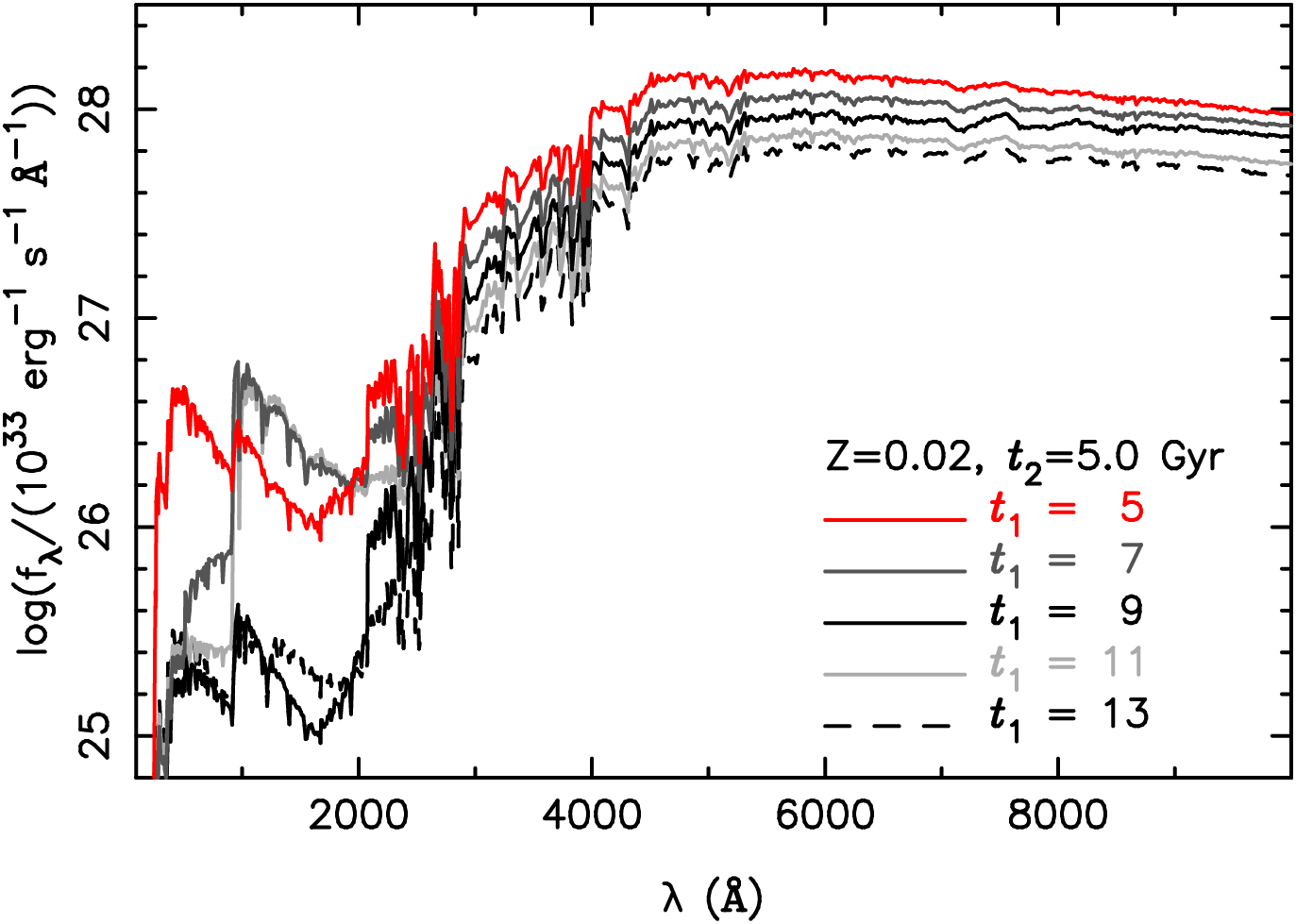}
\caption{Similar to Fig. 3, but for bsCSPs with a solar metallicity. Note that the meanings of colours and line types in this figure are different from those in Fig. 3.}
\end{figure}

\begin{figure} 
\centering
\includegraphics[angle=-90,width=0.35\textwidth]{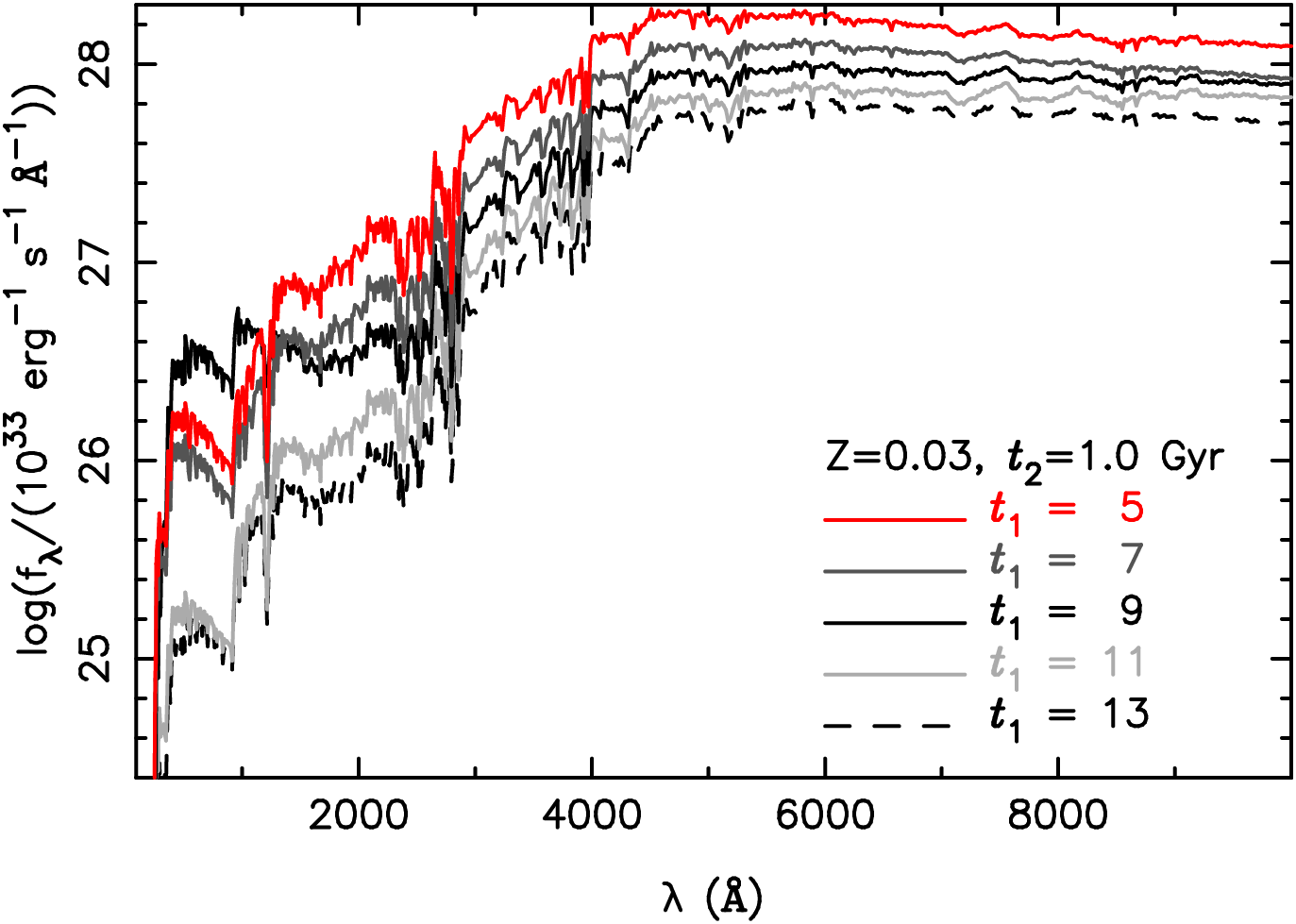}
\includegraphics[angle=-90,width=0.35\textwidth]{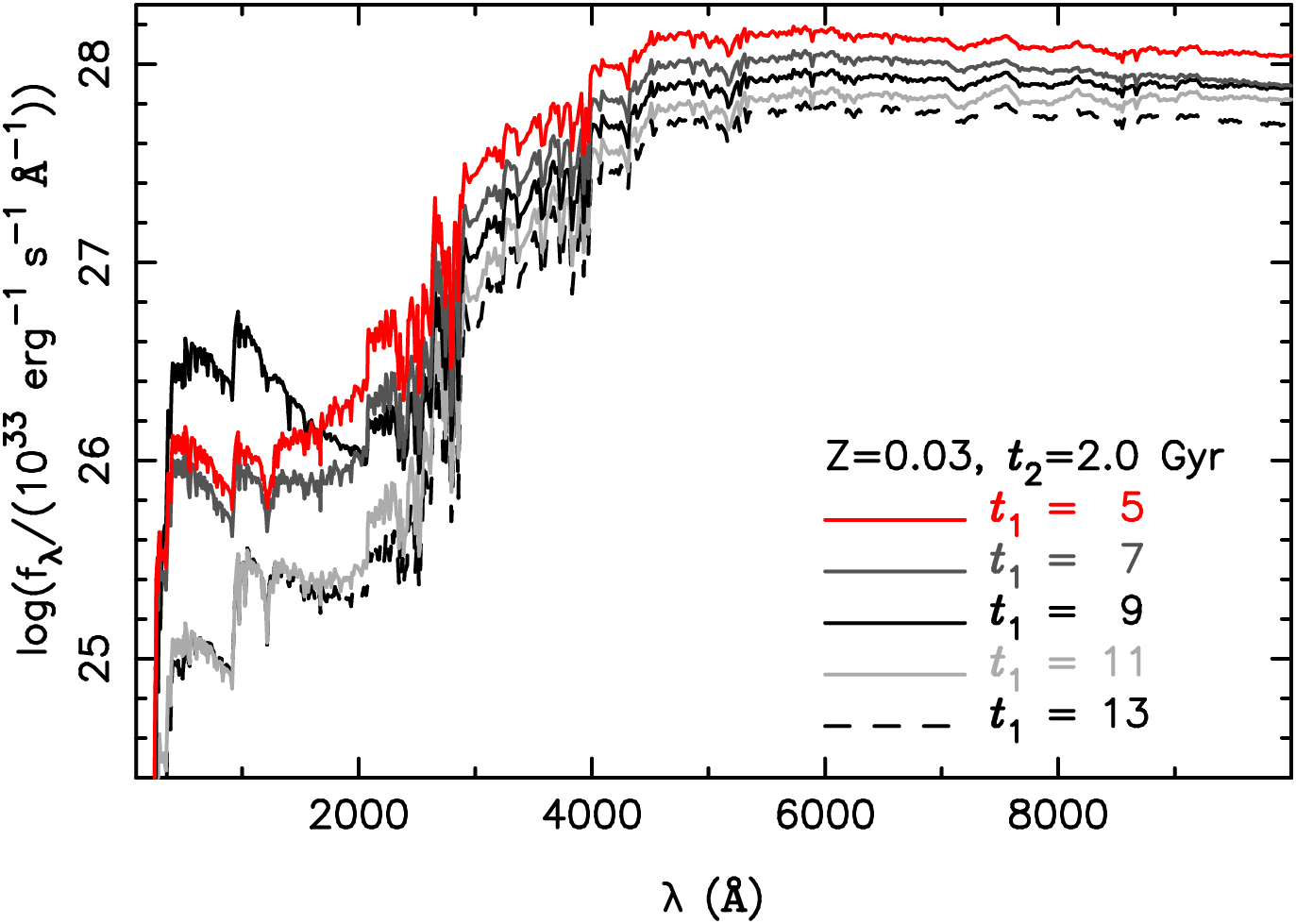}
\includegraphics[angle=-90,width=0.35\textwidth]{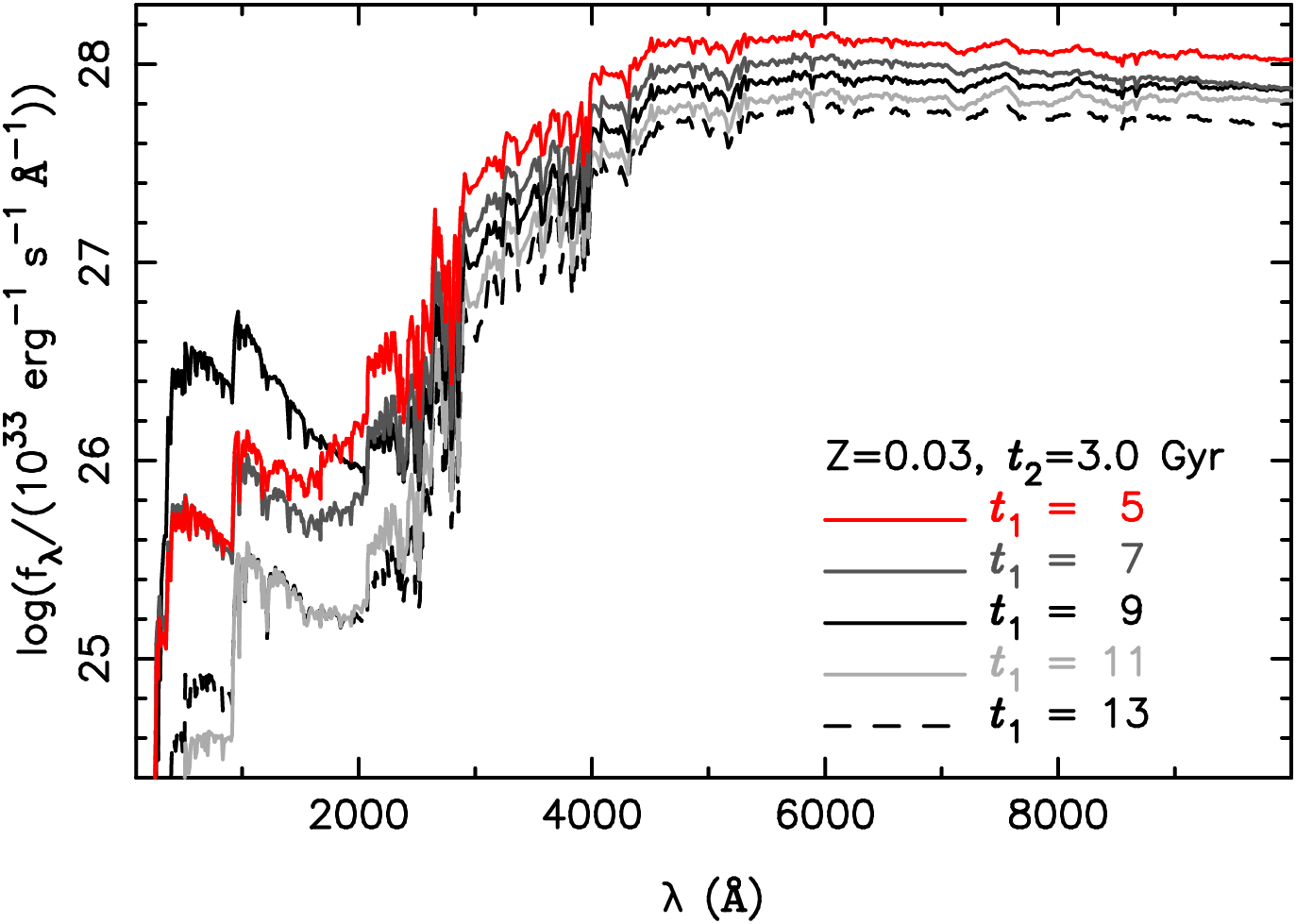}
\includegraphics[angle=-90,width=0.35\textwidth]{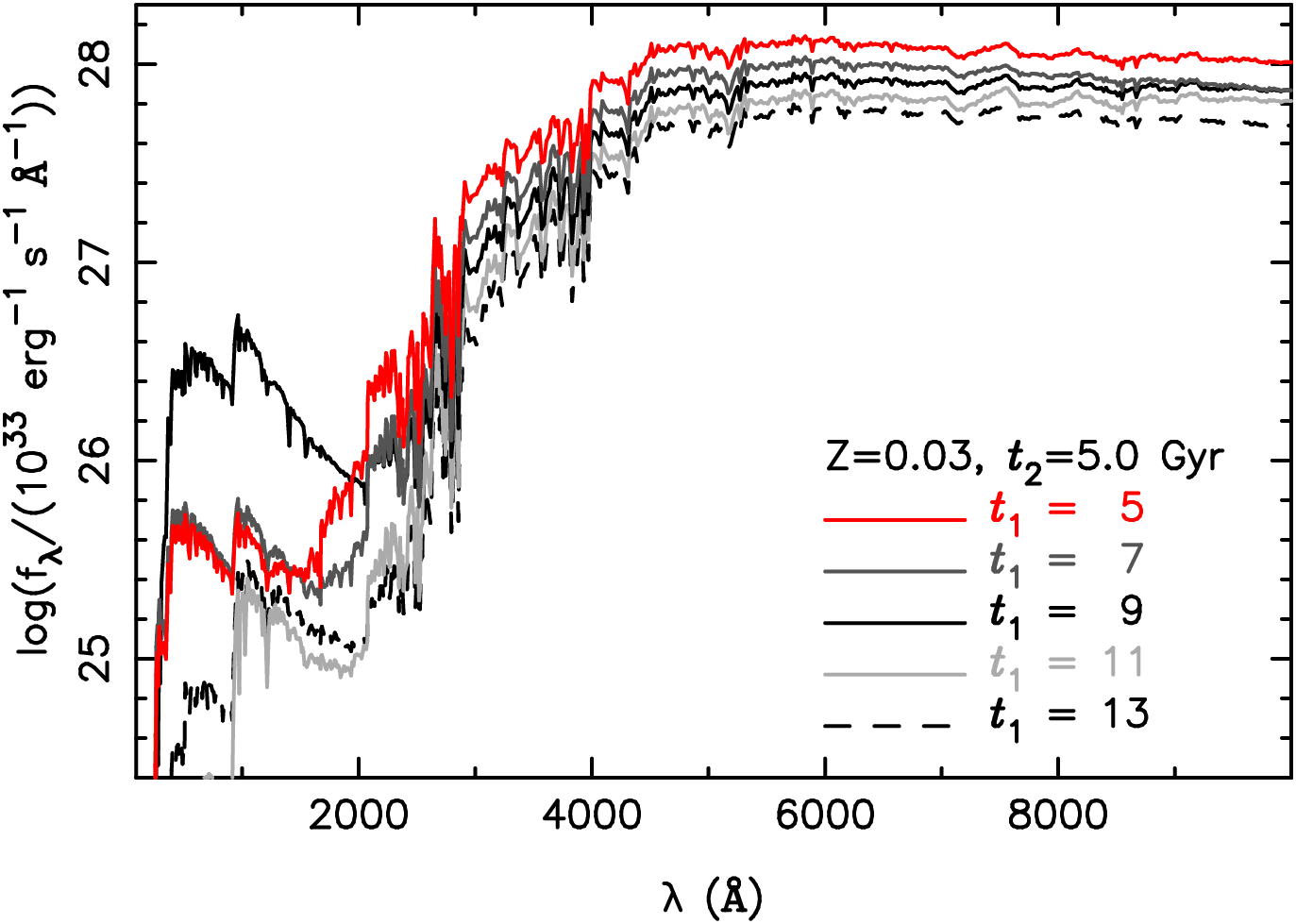}
\caption{Similar to Figure 3, but for bsCSPs with a metallicity ($Z$) of 0.03. Colours and line types have different meanings comparing to Fig. 3.}
\end{figure}

\subsection{Effect of young-component age}
The effects of the age of the young component, $t_{\rm 2}$, on the
SEDs of bsCSPs is investigated in this section. As examples, three
stellar metallicities (0.01, 0.02, and 0.03) and four ages (5, 8,
10, and 13\,Gyr) are taken for the stellar metallicity and
old-component age of bsCSPs, respectively. The results are shown in
Figs. 6 to 8. It is found that the age of young component has no
obvious effect on SEDs, when the young components of bsCSPs are
older than 2\,Gyr. However, $t_{\rm 2}$ affects the SEDs of bsCSPs
significantly when $t_{\rm 2}$ is less than 2\,Gyr. In detail,
the change of logarithmic flux of UV SED of bsCSP can be as large as
2\,dex, when $t_{\rm 2}$ changes in a wide range, e.g.,
0.5--10\,Gyr. In addition, it is found that the younger the young
component, the larger the flux in UV band, for bsCSPs with the same
metallicity ($Z$) and old-population age ($t_{\rm 1}$). In addition,
the three figures show that the shape of SED does not change obviously
for wavelength shorter than about 1000 $\rm \AA$, when $t_{\rm 2}$
changes from $t_{\rm 1}$ to 0.5\,Gyr. Meanwhile, the shape of SED
from 1000 to 3000 $\rm \AA$ is affected by $t_{\rm 2}$ more
obviously. This confirms that the SED within a wavelength range of
1000 to 3000 $\rm \AA$ is sensitive to the age of young component.
This agrees to many previous works, e.g., \citet{Yi:1999ApJ} and
\citet{Han:2007}.

\begin{figure} 
\centering
\includegraphics[angle=-90,width=0.35\textwidth]{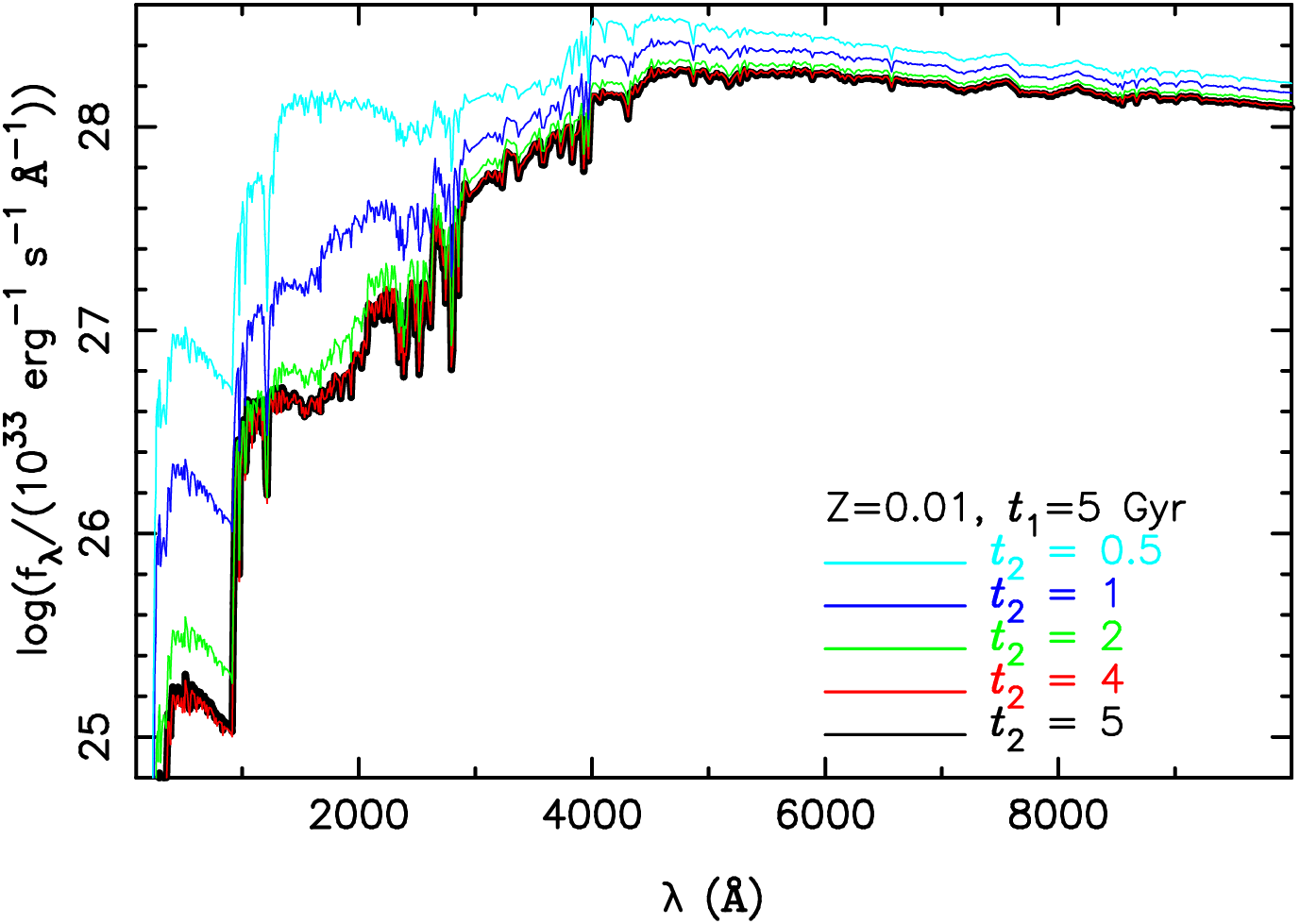}
\includegraphics[angle=-90,width=0.35\textwidth]{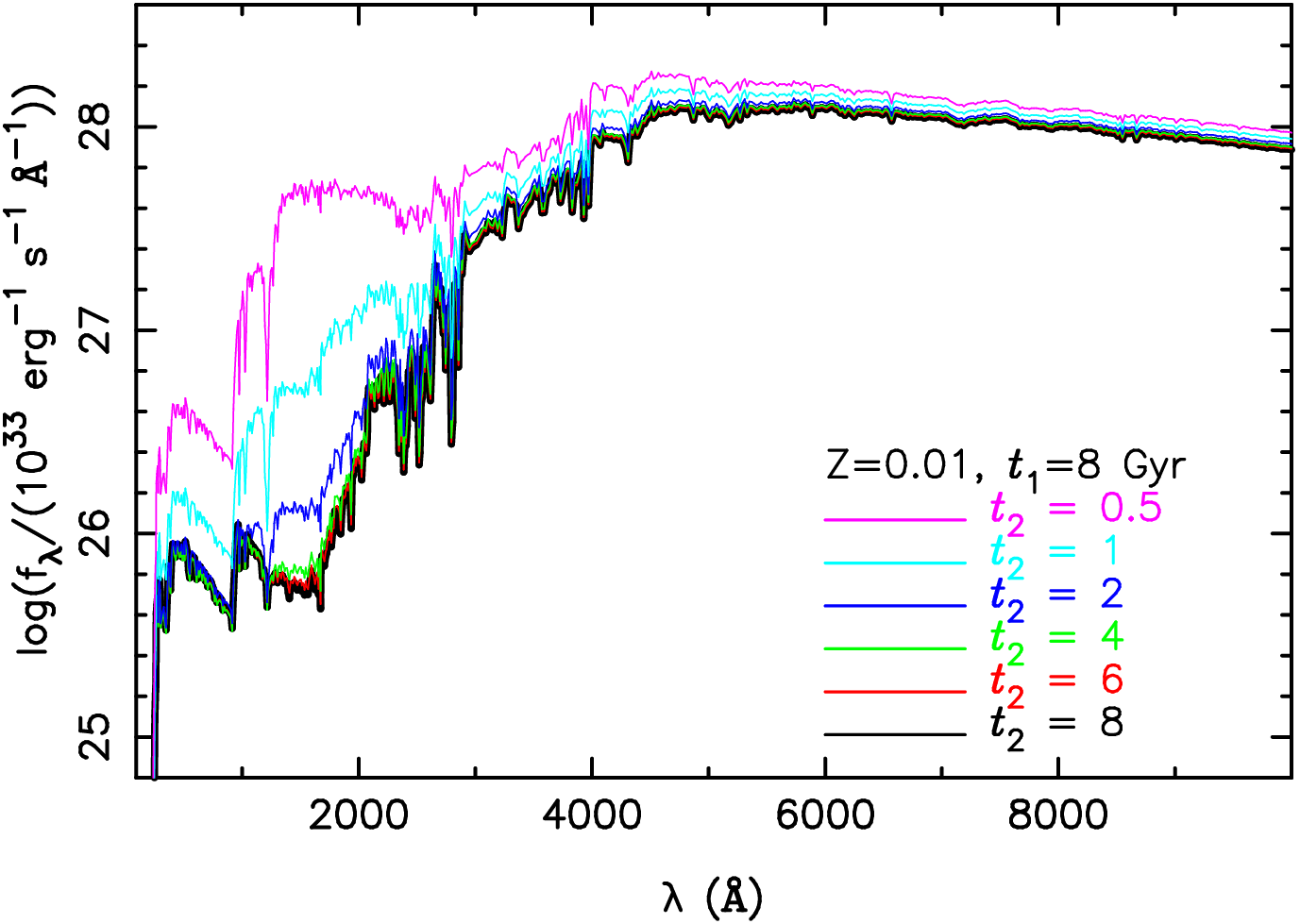}
\includegraphics[angle=-90,width=0.35\textwidth]{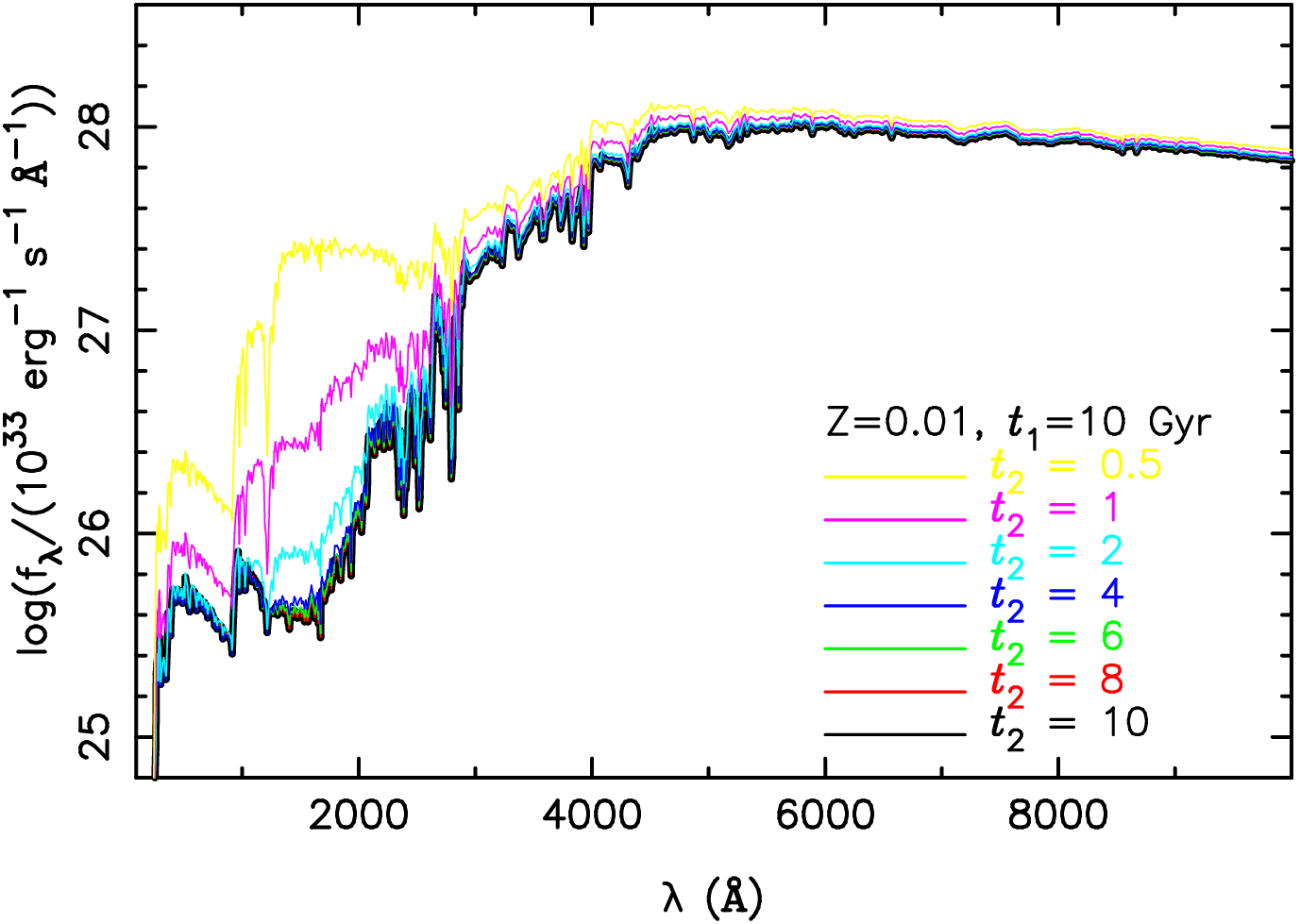}
\includegraphics[angle=-90,width=0.35\textwidth]{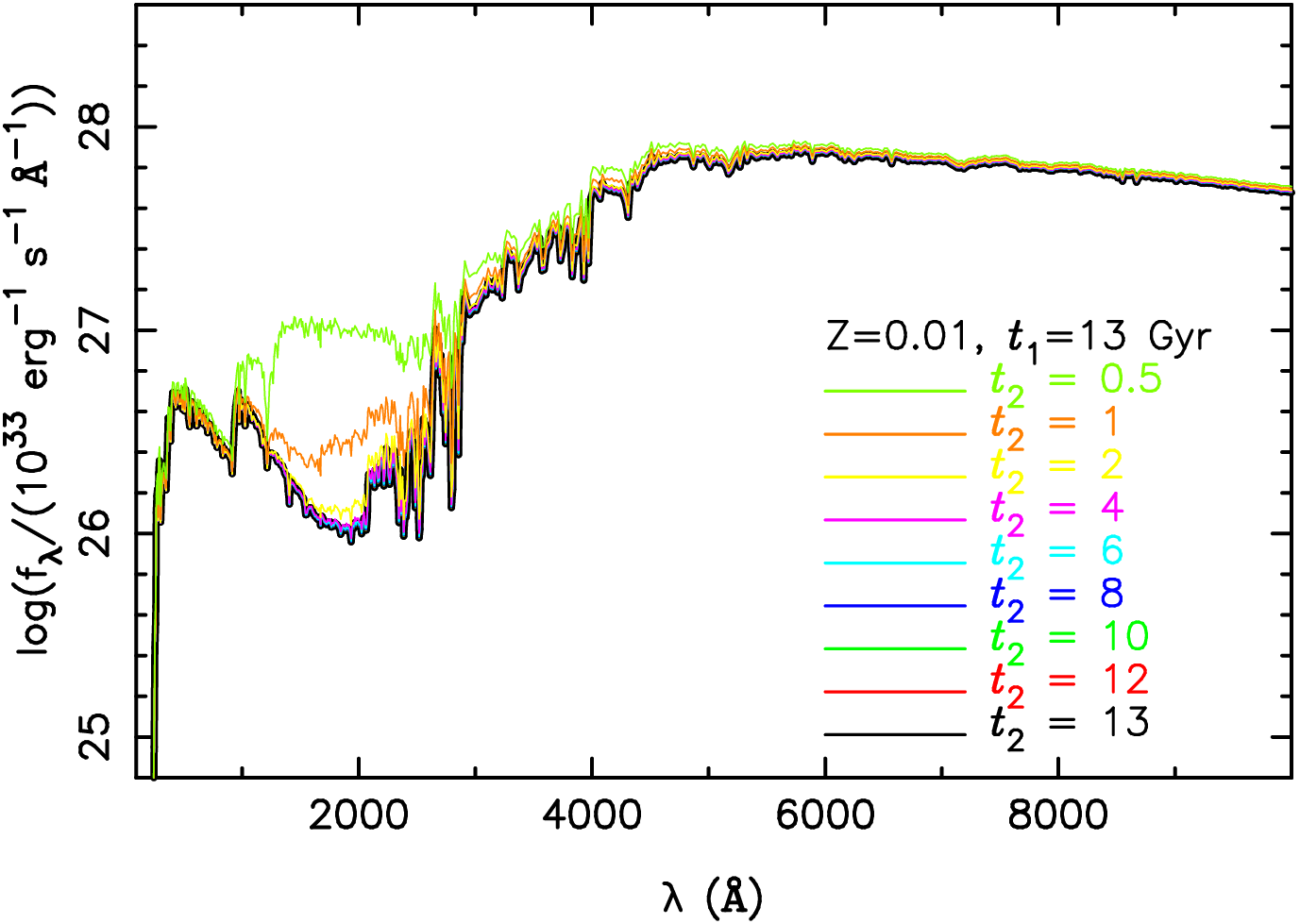}
\caption{Comparison of SEDs of bsCSPs with the same metallicity and old-component age, but various young-component ages.
In each panel, bsCSPs with various young-component ages are shown in different colours.
Note that colours have different meanings in four panels.}
\end{figure}

\begin{figure} 
\centering
\includegraphics[angle=-90,width=0.35\textwidth]{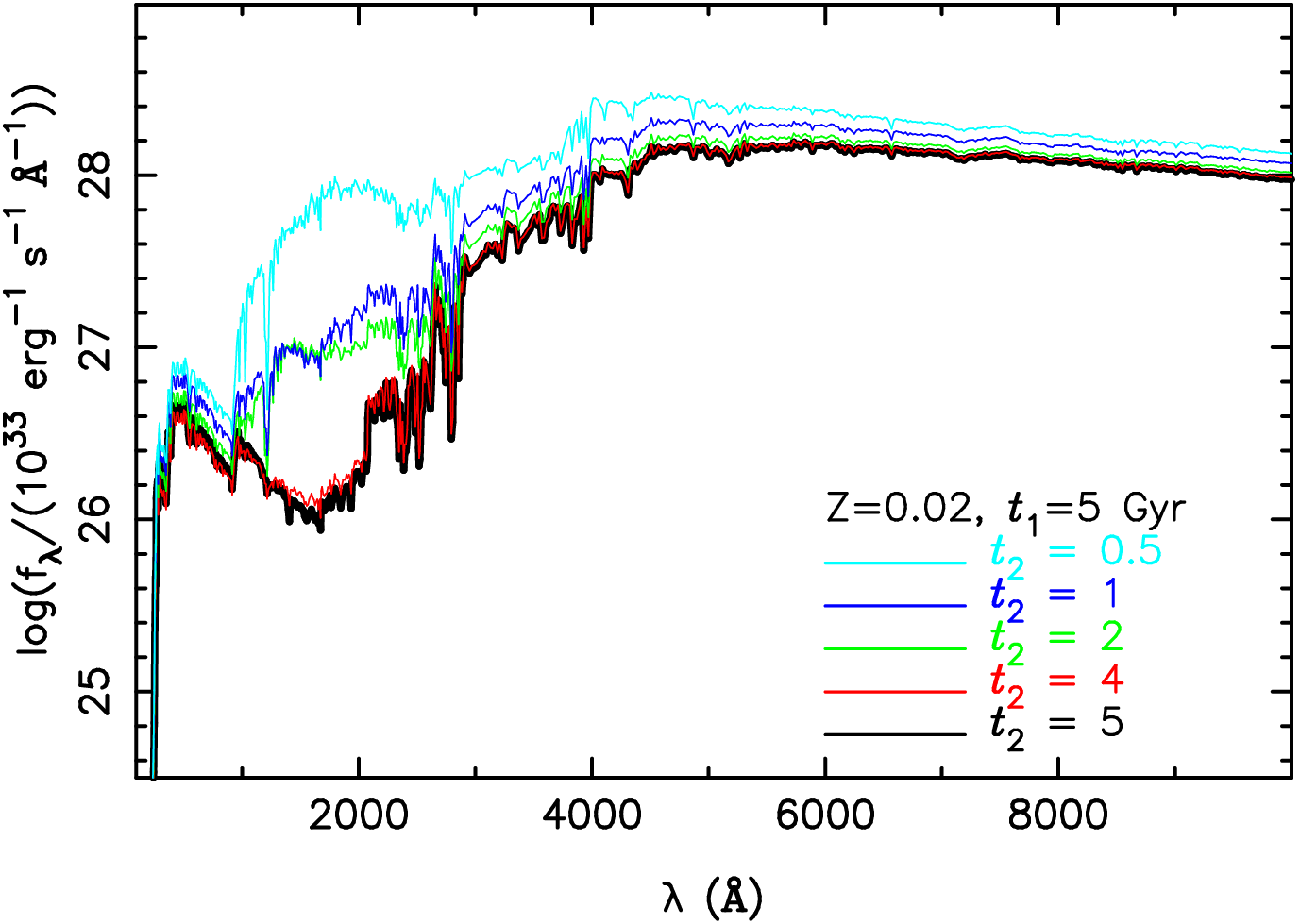}
\includegraphics[angle=-90,width=0.35\textwidth]{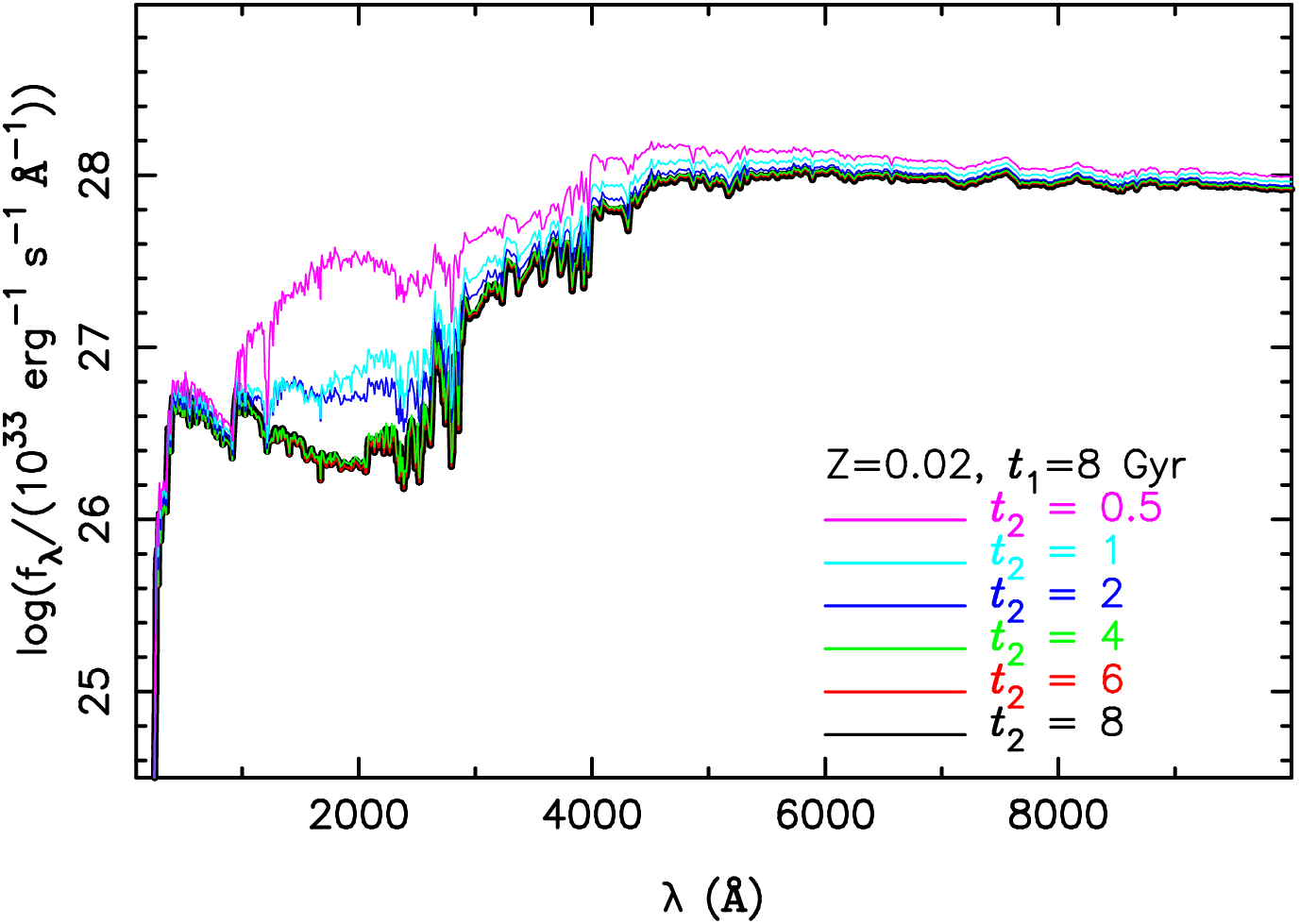}
\includegraphics[angle=-90,width=0.35\textwidth]{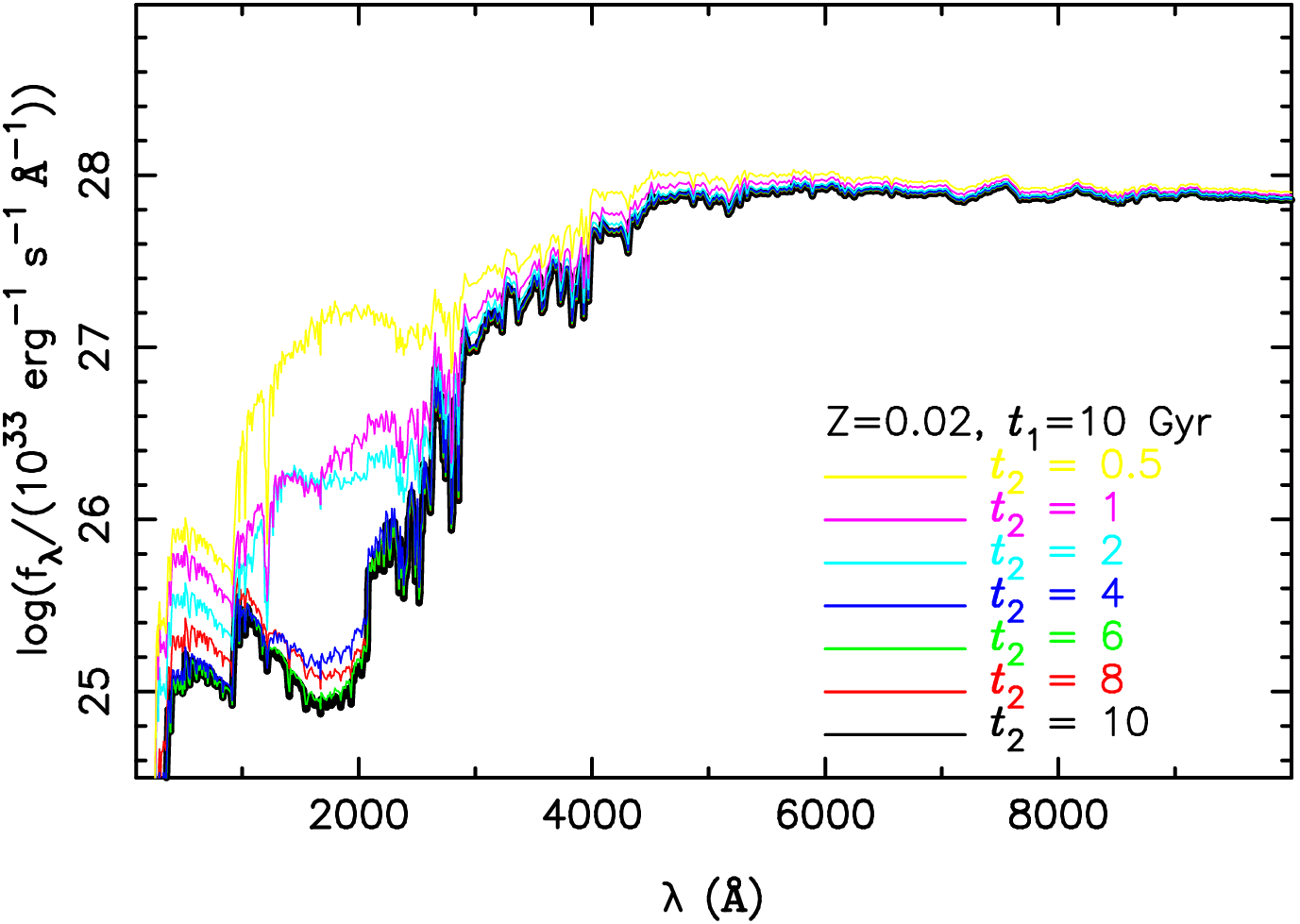}
\includegraphics[angle=-90,width=0.35\textwidth]{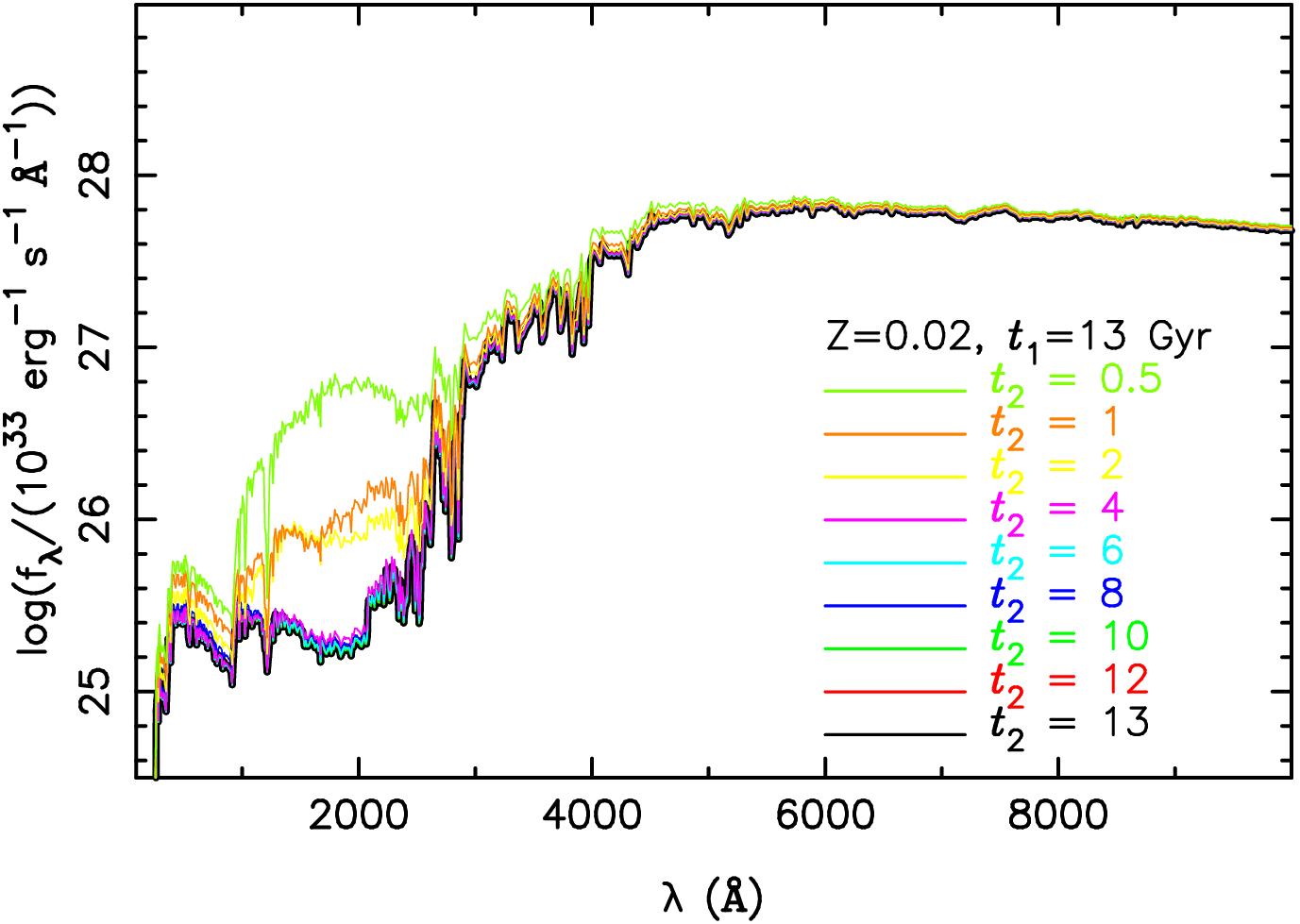}
\caption{Similar to Figure 6, but for bsCSPs with a solar metallicity ($Z$ = 0.02).}
\end{figure}

\begin{figure} 
\centering
\includegraphics[angle=-90,width=0.35\textwidth]{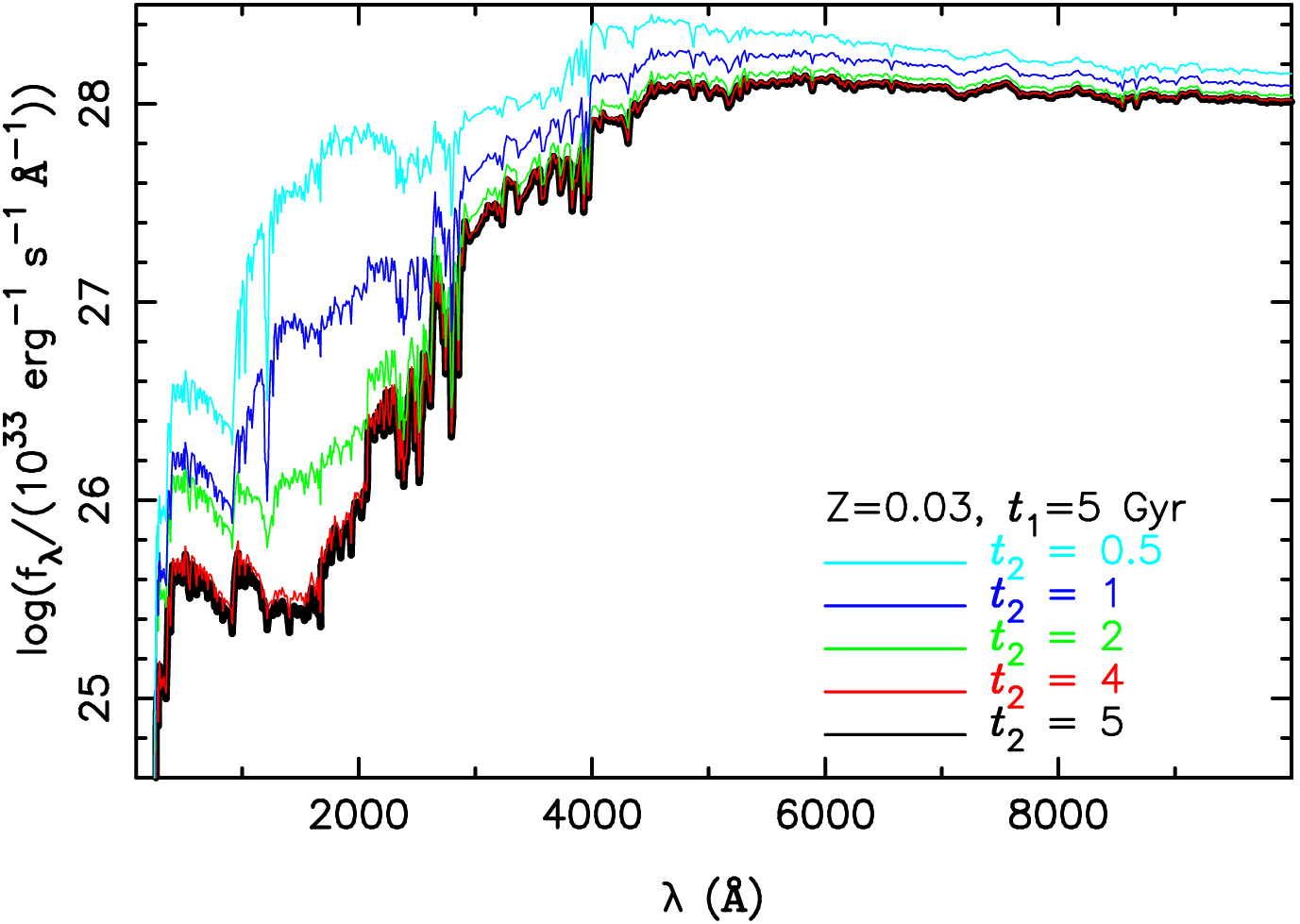}
\includegraphics[angle=-90,width=0.35\textwidth]{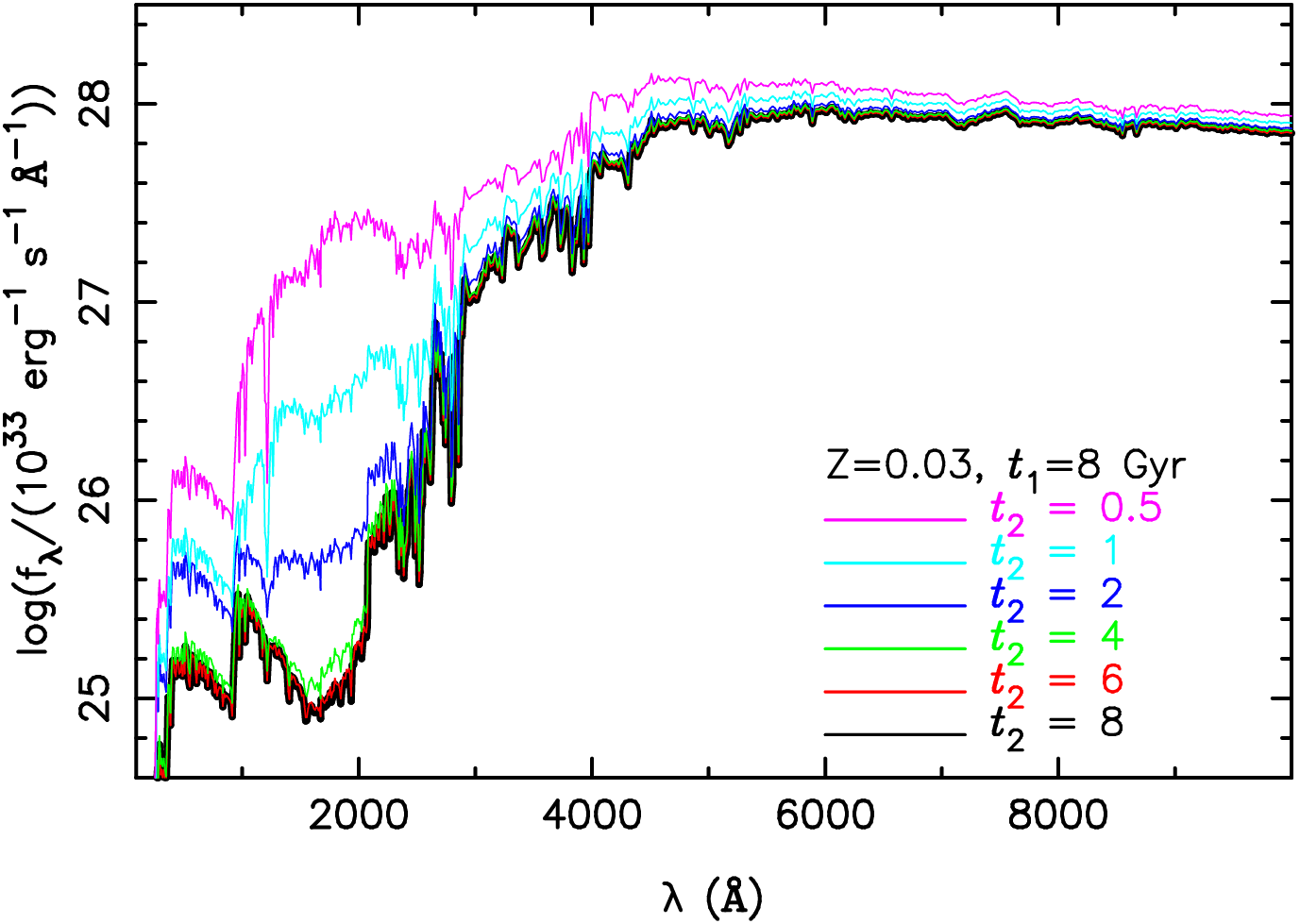}
\includegraphics[angle=-90,width=0.35\textwidth]{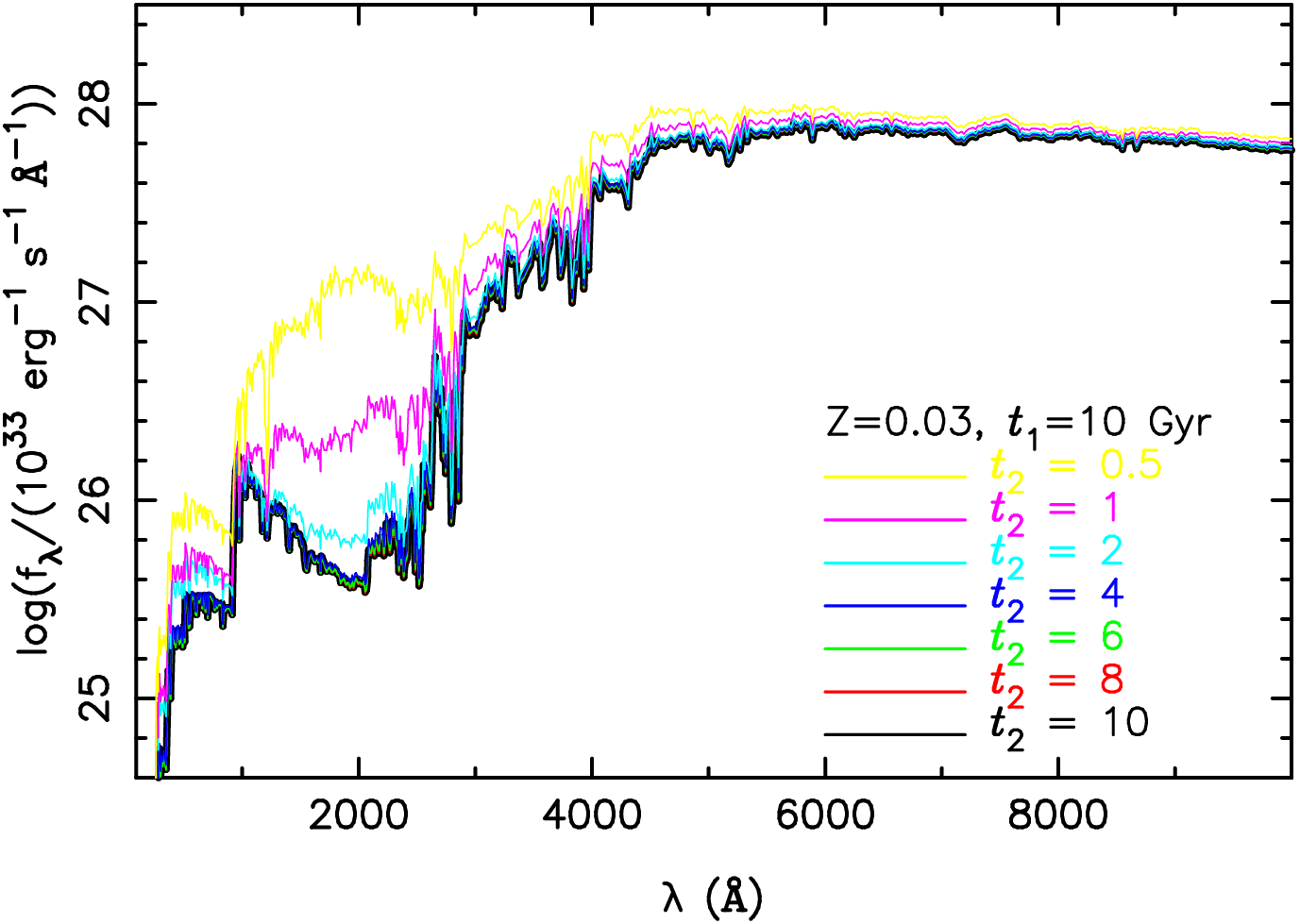}
\includegraphics[angle=-90,width=0.35\textwidth]{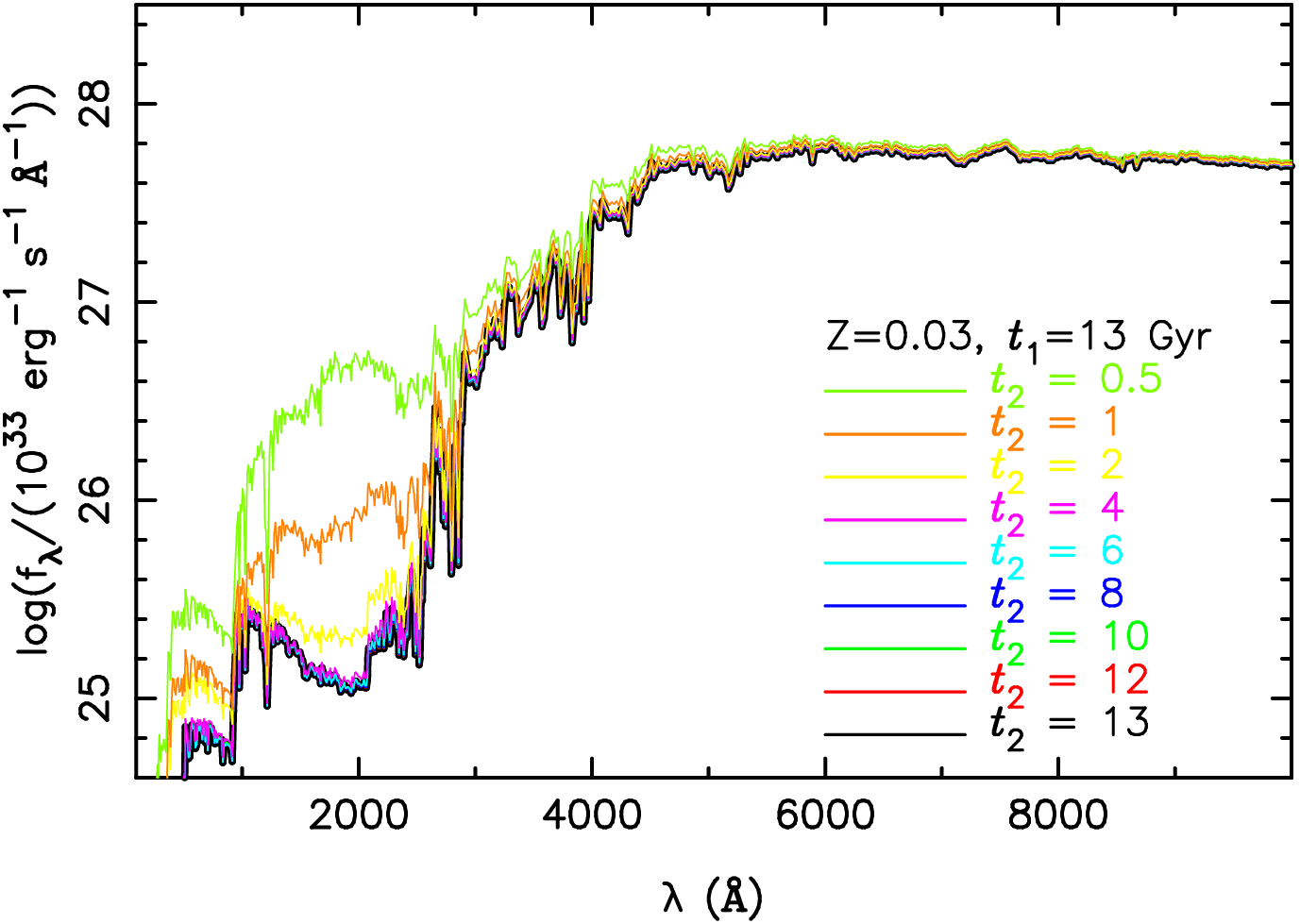}
\caption{Similar to Figure 6, but for bsCSPs with a metallicity ($Z$) of 0.03.}
\end{figure}

\section{Sensitivities of SEDs to stellar population model inputs}
Because three input stellar population parameters (metallicity, old
and young-component ages) have some effects on the SEDs of bsCSPs,
it is difficult to find SED features for determining each parameter
from the results shown in Section 3. We investigate the
sensitivities of SEDs to three main input parameters of bsCSPs. The
absolute sensitivity to an input stellar population parameter is
defined as the average change of logarithmic flux (log $f_{\rm
\lambda}$) at a wavelength, when the input parameter changes by 10\%
(0.0026 for $Z$, 0.7\,Gyr for old-component age, and 0.6\,Gyr for
young-component age) within its possible range. The absolute
sensitivities are calculated using the SEDs of bsCSPs with 4
metallicities (0.004, 0.01, 0.02, and 0.03) and 8 old-component ages
(from 5 to 12\,Gyr with a 1\,Gyr interval). The ages of young
components of bsCSPs take different values from its old-component
age to 1\,Gyr. Note that because this work aims to study the SEDs of
early-type galaxies, the possible ranges for the metallicities and
old-component ages are taken from a result of stellar population
study of nearby early-type galaxies (e.g., \citealt{Gallazzi:2005}).

The absolute sensitivities at all wavelengths are shown in Fig. 9.
It is found that the sensitivities to three stellar population
parameters differ a lot at various wavelengths. As a whole, SEDs in
near infrared and infrared bands ($\lambda \geq$ 8000\,$\rm \AA$)
are more sensitive to old-component age, while UV band SED is
sensitive to all of three stellar population parameters. In other
words, old-component age affects the brightness in near infrared and
infrared bands, significantly. The brightness in UV band is affected
by all of three stellar population parameters. As we see, UV flux is
affected by stellar metallicity significantly. The optical SEDs of
bsCSPs are less sensitive to the age of young components, compared
to the stellar metallicity and age of old components of bsCSPs. This
suggests that it is difficult to determine the young-component ages
of bsCSPs via optical SEDs. Therefore, the results suggest to use
multiple-band SEDs for well determining the stellar metallicity,
old-component and young-component ages of early-type galaxies, via
fitting SEDs.

Although Fig. 9 tells us the absolute changes in SEDs caused from 10
percent change of stellar metallicity, old-component age, or
young-component age, it is difficult to guide many of our studies.
The reason is that usually we determine the stellar population
parameters via fitting some relative spectral features (or shapes)
of galaxies. Such special spectral features can be found by
searching for the abrupt changes with wavelength of the relative
sensitivities of SEDs to three stellar population parameters. For
convenience, we use relative sensitivities instead of absolute
sensitivities here. The relative sensitivities of SEDs to stellar
metallicity is obtained by dividing absolute sensitivities to
metallicity by 1.2. The relative sensitivities of SEDs to
old-component age and young-component age are obtained by dividing
absolute sensitivities by 0.7. This makes the maximum of relative
sensitivities of SEDs to three stellar population parameters equal
to 1 for $\lambda >$ 500\,$\rm \AA$ . Then some spectral features
can be found from the relative sensitivities. For example, if the
sensitivity to old-component age within a special wavelength range
is significantly larger than in neighborhood, while the
sensitivities to stellar metallicity and young-component age in the
same wavelength range do not have similar trend, the SED feature
within the given wavelength range is relatively sensitive to
old-component age. It means the relative shape of SEDs within the
given wavelength range can be used as an old-component age
indicator. One can check the sensitivities of SEDs within
1205--1245\,$\rm \AA$ (peak at 1225\,$\rm \AA$) in Fig. 10 for well
understanding our method. Some spectral features that are sensitive
to stellar metallicity and young-component age are found in the same
way. Fig. 10 shows the relative sensitivities of SEDs to three
stellar population parameters and the central wavelengths of some
example spectral features that are sensitive to stellar metallicity,
old-component age, or young-component age of bsCSPs. In detail, the
results show that SED features within wavelength ranges of
5110-5250\,$\rm \AA$, 5250--5310\,$\rm \AA$, 5310--5350\,$\rm \AA$,
5830--5970\,$\rm \AA$, 20950--23550\,$\rm \AA$ are relatively
sensitive to stellar metallicity of bsCSPs. The SED features within
965-985\,$\rm \AA$, 1005--1055\,$\rm \AA$, 1205--1245\,$\rm \AA$ are
sensitive to old-component age, while those within the wavelength
ranges of 2185--2245\,$\rm \AA$, 2455--2505\,$\rm \AA$,
2505--2555\,$\rm \AA$, 2775--2825\,$\rm \AA$, 2825--2875\,$\rm \AA$
to young-component age.

It is found that although the UV SEDs of bsCSPs are related to all
of three stellar population parameters, some spectral features
(e.g., 1205 $\leq \lambda \leq$ 1245\,$\rm \AA$ ) in UV band are
found to be  more sensitive to the old-component age. It is also
found from Fig. 10 that SED features within 5110-5250\,$\rm \AA$,
5250--5310\,$\rm \AA$, 5310--5350\,$\rm \AA$ are relatively
sensitive to stellar metallicity. This suggests that even bsCSP
models instead of ssSSP models are used for studying the stellar
populations of early-type galaxies, some widely used spectral line
indices, e.g., Mg$_{\rm b}$ Fe5270, Fe5335, can be used as
metallicity indicators. In addition, SED features in a wide
wavelength range of near infrared band seem sensitive to stellar
metallicity. It implies that infrared photometry has some potential
for determining the stellar metallicity of galaxies. Moreover, it is
shown that the age of young-components of bsCSPs of early-type
galaxies can be estimated by some SED features within narrow
wavelength ranges, e.g., 2775--2825\,$\rm \AA$ (with a peak
wavelength of 2800 $\rm \AA$), because SEDs in these wavelength
ranges are relative sensitive to young-component age of bsCSPs.

\begin{figure} 
\centering
\includegraphics[angle=-90,width=0.4\textwidth]{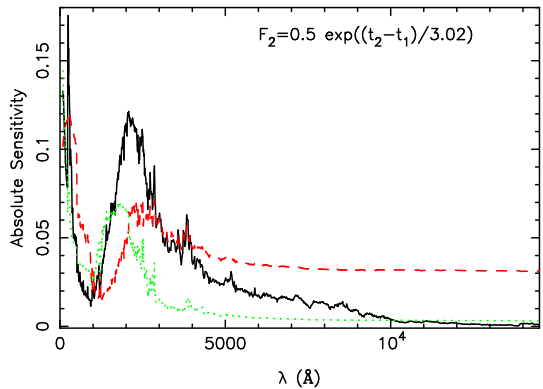}
\caption{Absolute sensitivities of different SED parts to three stellar population parameters of bsCSPs.
Black solid, red dashed, and green dotted lines show the sensitivities to metallicity, old-component age, and young-component age, respectively.}
\end{figure}

\begin{figure} 
\centering
\includegraphics[angle=-90,width=0.48\textwidth]{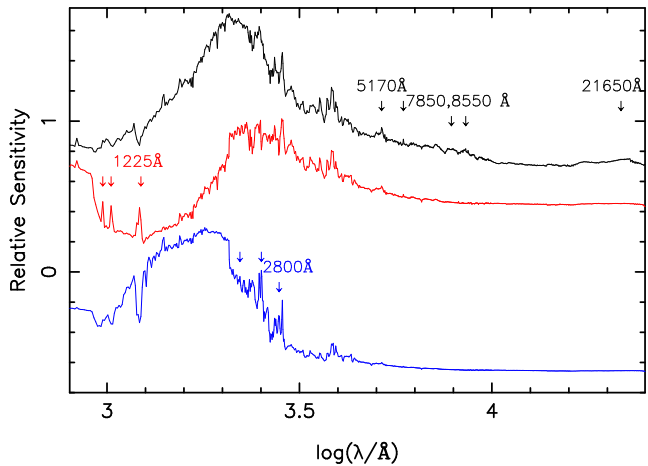}
\caption{Relative sensitivities of different SED parts to three stellar population parameters of bsCSPs.
Relative sensitivities to metallicity, and young-component age are added by 0.7 and -0.7 to make them separate from the sensitivities to old-component age.
Black (top), red (mid), and blue (bottom) lines show the sensitivities to metallicity, old-component age, and young-component age, respectively.}
\end{figure}

\section{Application and comparison of bsCSP models to galaxies}
As a test, in this section we apply the above results, i.e., spectra
database of bsCSPs and sensitivities of different spectral parts to
the input stellar population parameters, to three elliptical
galaxies, i.e., NGC1399, NGC1404, and NGC2865. Stellar population
parameters of the three galaxies are determined via spectral fits.
The observational spectra of three galaxies are taken from a
database of UV-optical spectra of nearby quiescent and active
galaxies, which was built by Storchi-Bergmann, Calzetti \& Kinney
(\url{http://www.stsci.edu/ftp/catalogs/nearby_gal/sed.html}),
because of its convenience. Then the original spectra were degraded
to fit the resolution of BaSeL 3.1 spectra. Some bsCSP models with
metallicities of 0.0003, 0.001, 0.004, 0.01, 0.02 and 0.03 are used
for fitting the observational spectra, and all spectra are
normalized at 5510 $\rm \AA$. The range of old-population age of
bsCSP models is 0.1--15\,Gyr. Two methods are used to get the best
fit results. In the first way, a homogeneous weight is given for
almost all spectral parts. The spectral parts that have obvious
uncertainties or poor quality in the observational spectra are given
lower weights in all fits. The results show that NGC1399, NGC1404,
NGC2865 have redshifts of 0.004753, 0.00650 and 0.008797,
respectively. Their best fit stellar metallicities are shown to be
0.03, 0.03, and 0.01, respectively. Meanwhile, the best fit ages for
the old and young populations in NGC1399, NGC1404 and NGC2865 are
(11.8 and 10.9 Gyr), (10.3 and 1.7 Gyr) and (14.7 and 1.1Gyr),
respectively. When various weights are given to different spectral
parts in the second way, the best fit stellar populations of NGC1399
and NGC1404 are shown to be different. The metallicity of NGC1404 is
shown to be 0.02, instead of 0.03. NGC1399 is shown to include only
an old (11.8\,Gyr) population, while NGC1404 is shown to include a
pair of populations with ages of 13.8 and 1.3\,Gyr. The detailed
comparison of the observed and best fit bsCSP spectra of the
galaxies are shown in Figs. 11 and 12. Note that in the second way,
four times weights are given to three spectral parts, i.e.,
1205--1245$\rm \AA$, 2775--2885$\rm \AA$, 5110--5350$\rm \AA$,
according to their relative sensitivities to stellar population
inputs. We see that the observational spectra of three elliptical
galaxies can be fitted well using those of bsCSPs, as a whole.
However, it is also seen that the spectral parts around 2000 and
3000$\rm \AA$ are not fitted well. This actually results from the
poor quality of the observational spectra around 2000 and 3000$\rm
\AA$, because the part around 2000$\rm \AA$ is a overlap region of
two observational spectra and the second order contamination of
light causes uncertainties in the part of 3000--3200$\rm \AA$. In
addition, only bsCSPs with six metallicities are use for the fit.
This may also contribute to the differences between the
observational and best fit spectra.

\begin{figure} 
\centering
\includegraphics[angle=-90,width=0.47\textwidth]{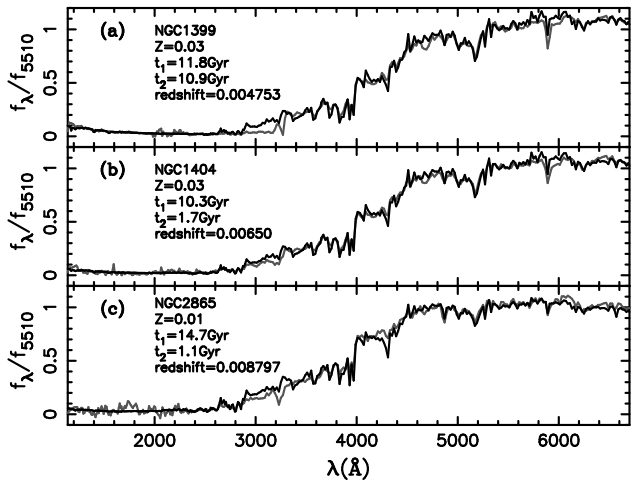}
\caption{Comparison of observational and bsCSP fit spectra of three elliptical galaxies.
All spectra are normalized at 5510 $\rm \AA$. Gray and black lines are for observational and theoretical spectra,
respectively. The best fit redshifts and stellar population parameters can be seen in three panels. Homogeneous
weight is given to all spectral parts in spectral fit. Note that the parts around 2000 and 3000 $\rm \AA$ of
observational spectra contains larger uncertainties.}
\end{figure}

\begin{figure} 
\centering
\includegraphics[angle=-90,width=0.47\textwidth]{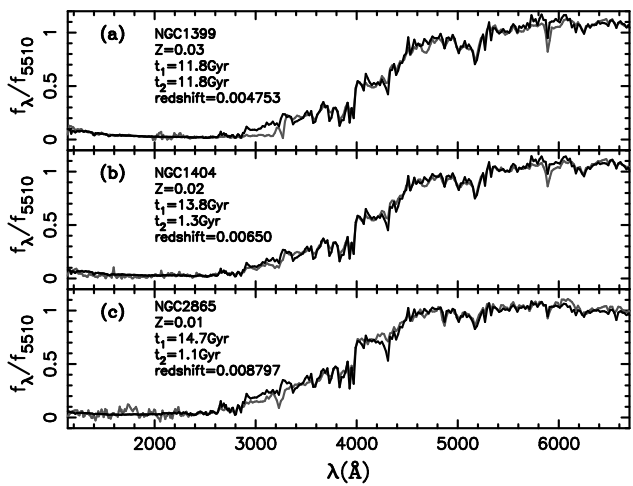}
\caption{Similar to Fig. 11, but various weights are given to different spectral parts in the spectra fit
according to the different relative sensitivities of various parts.}
\end{figure}

\section{Conclusion and discussion}
Aiming to study the SEDs of early-type galaxies, this paper presents
the SEDs of binary star composite stellar populations (bsCSPs) and
studies the sensitivities of different parts of SEDs of bsCSPs to
stellar metallicity, old-component and young-component ages. Some
SEDs with UV-upturn are presented naturally for some bsCSPs. Our
results also show detailed effects of three stellar population
parameters on the SEDs of bsCSPs. It suggests that all of the three
stellar population parameters should be taken into account in
studies of SEDs of early-type galaxies.

In particular, this work shows clear effects of stellar metallicity
on the UV SEDs of bsCSPs, even when metallicity changes within a
range near the solar value, e.g., 0.01--0.03. Therefore, our results
suggest to use bsCSPs with different metallicities for studying the
SEDs of early-type galaxies, including studying UV-band SEDs. In
addition, the results show that the flux in optical and infrared
bands are dominated by the old-component age of bsCSPs, while the
flux in UV band is affected by all of three stellar population
parameters. However, it seems that metallicity determines the
UV-band SEDs of bsCSPs, because it is much more sensitive to UV SED
compared to the ages of two stellar population components. Our
results suggest using multi-band SEDs to determine the sellar
metallicity, old-component age, and young-component age of bsCSPs.

This work also showed some special wavelength ranges for studying
bsCSPs of early-type galaxies. The results are potentially useful
for determining the sellar metallicity, old-component age, and
young-component age of bsCSPs of early-type galaxies via fitting
relative spectral features (e.g., spectral line indices) of SEDs.

As a standard model, we take a formulae of $F_{\rm 2} = 0.5 ~{\rm
exp}[(t_{\rm 2}-t_{\rm 1})/{\rm \tau}]$ with $\tau$ = 3.02 for
calculating the mass fraction of the young component of a bsCSP. The
value of $\tau$ affects the sensitivities of SED to the
young-component age of bsCSPs obviously, because it changes the mass
fractions of young-components of bsCSPs directly. We checked the
sensitivities of SED to bsCSPs by taking different values (1, 2,
3.02 and 5) for $\tau$. The results showed that the bigger the value
of $\tau$, the larger the sensitivity of UV-band SEDs to
young-component age. However, the sensitivities of optical and
infrared-band SEDs to the young-component age are always similar.
The sensitivities of SEDs to metallicity and old-component age do
not change obviously with different $\tau$ values.

The initial mass function (IMF) of stellar populations can also
affect the sensitivities of SED to stellar metallicity,
old-component and young-component ages. However, even other IMFs are
taken for building bsCSPs, the main results of this work will not
change. This is checked using some bsCSPs with a
\citet{Salpeter:1955} IMF. Therefore, the results of this paper can
be used widely for stellar population studies. One can also refer to
some works such as
\citet{Conroy:2009ApJ699,Conroy:2010ApJ708,Conroy:2010ApJ712} for
better understanding the uncertainties in evolutionary population
synthesis.

Because the stellar metallicity of most nearby early-type galaxies
are shown to be higher than 0.004, we did not show the results for
bsCSPs with lower metallicity. However, our study showed that if
lower metallicities are taken for bsCSPs, their SEDs will change
significantly. The effects of $\alpha$-element enhancement was not
taken into account in this work, because the Hurley code can not
evolve stars with $\alpha$-element enhancement. One can read recent
papers, e.g., \citet{Lee:2009AJ} for reference.

\section*{Acknowledgments}
We thank Profs. ZHAO Gang and SHI Jianrong for useful suggestions.
We also thank Prof. Vanbeveren for comments and revision
suggestions. This work has been supported by the Chinese National
Science Foundation (Grant Nos. 10963001, No. 10978010), Yunnan
Science Foundation (No. 2009CD093), and Chinese Postdoctoral Science
Foundation. Zhongmu Li gratefully acknowledges the support of
Sino-German Center (GZ585) and K. C. Wong  Education Foundation,
Hong Kong. Liu's work was also supported by the light in China's
Western Region (LCWR)(No.XBBS2011022).


\label{lastpage}

\end{document}